\documentclass{eas}
\usepackage{graphicx}
\newcommand{\mdot}{\ensuremath{\dot{M}}}                             

\begin{document}

\title{Nucleosynthesis from massive stars 50 years after B$^2$FH} 
\runningtitle{Rotating massive stars}
\author{Georges Meynet}\address{Geneva Observatory, University of Geneva, CH-1290 Sauverny, Switzerland}
\begin{abstract}
We review some important observed properties of massive stars. Then we discuss how mass loss and rotation affect their evolution and help in giving
better fits to observational constraints. Consequences for nucleosynthesis at different metallicities are discussed. 

Mass loss appear to be the key feature at high metallicity, while rotation is likely dominant at low and very low metallicities. We discuss various indications supporting the view that
very metal poor stars had their evolution 
strongly affected by rotational mixing. Many features, like
the origin of primary nitrogen at low metallicity, that of the C-rich extremely metal poor halo stars, of He-rich stars in massive globular clusters, of the O-Na anticorrelation in globular clusters may be related to the existence of a population of very fast rotating metal poor stars that we tentatively call the {\it spinstars}.
A fraction of these {\it spinstars} may also be the progenitors of GRB in very metal poor regions. They may avoid pair instability explosion due to the heavy mass loss undergone during their early evolutionary phases and be,
dependent on their frequency, important sources of ionising photons in the early Universe.
\end{abstract}
\maketitle
\section{From stars to nuclei, what's new after the B$^2$FH paper?}

The papers by Burbidge, Burbidge, Fowler and Hoyle (1957, B$^2$FH ) and Cameron (1957) are wonderful achievements exposing in a detailed and
documented manner nearly all the nuclear processes responsible for nucleosynthesis in the Universe. In the last fifty years, only very few new nuclear processes have been added to those listed in these papers. This
underlines the success of the authors in providing an exhaustive and correct synthesis still accepted today (a very nice account of the history of the field of nucleosynthesis has been written in french by Celnikier 1998). Among the new processes described later, let us mention the spallation reactions resulting from the interactions of the cosmic rays with the interstellar medium and responsible for the synthesis of lithium (at least of a part of it), beryllium and bore (Reeves et al. 1970; see also the x-process in B$^2$FH),  the neutrino processes (Domogatsky \& Nadyozhin 1977, Nadyozhin 1991) and, more recently, the possible ``$\nu$p-process'' (Fr\"ohlich et al. 2006) believed to occur during core collapse supernovae. Does this mean that the chapter of stellar nucleosynthesis is close to its end, and does no long constitute an interesting field for topical research? It does not appear so, in contrary stellar nucleosynthesis appears at the heart of many topical problems going from stellar physics to that of the first stellar generations in the Universe, passing through the evolution of galaxies. Let us give some arguments supporting this view:
\begin{itemize}
\item Well known nuclear processes can be used to check interior stellar models, in particular mixing processes which play a key role in governing the photometric evolution of stars, their lifetimes (important for age determination of stellar systems), and nucleosynthesis. 
\item The abundances in the gas of galaxies is a powerful tool to discriminate between different 
star formation histories (see e.g. Matteucci 2001). The interpretation of these chemical tracers
does require a quantitative knowledge of stellar nucleosynthesis.
\item Abundances at the surface of very metal poor halo stars and in the gas of high redshift galaxies open new windows on the early phases of the chemical evolution of galaxies and of globular clusters. Many recent observations challenge the current views on the evolution of stars in metal poor regions (see e.g. Meynet et al. 2005).
The first generations of stars might have had peculiar properties whose consequences go beyond nucleosynthesis and affect their ionizing power, the nature of the supernova events and of the stellar remnants. 
\end{itemize}
The few examples above show that stellar nucleosynthesis has become a tool to investigate key evolutionary processes in the Universe. To go ahead in this direction it is however necessary to obtain reliable quantitative results for stellar nucleosynthesis and in that respect still a lot of work needs to be done. Two main lines of research have to be pursued simultaneously, on one hand fundamental physical processes occurring in stars have to be improved, on the other hand detailed and careful comparisons of models with observed characteristics of stars are necessary to continuously check the models. Let us recall here this sentence by Martin Schwarzschild (1958) in the prologue of his book ``Structure and evolution of the stars'':``{\it If simple perfect laws uniquely rule the universe, should not pure thoughts be capable of uncovering this perfect set of laws without having to lean on the crutches of tediously assembled observations?... In fact, we have missed few chances to err until new data freshly gleaned from nature set us right again for the next steps.}''. 

In the following we begin by recalling recent observations of massive stars that stellar models should be able to explain and reproduce quantitatively. In a second part, we present some improvements in the physics of massive star models. We shall focus here mainly on the interactions between rotation and mass loss, other processes being presented elsewhere in this volume (see the papers by Talon and Arnould). Then we present some comparisons between stellar models
and observations. In the last part we discuss some possible consequences for nucleosynthesis. We shall mainly focus here of the first phases of the evolution of massive rotating stars, the advanced  evolutionary phases as well as the explosive nucleosynthesis is treated by Limongi in the present volume.

\section{Some general properties of massive stars}

Let us imagine a universe without massive stars. The chemical evolution would be much slowlier and our universe would be a much less fascinating place (with probably nobody to make this assessment!). Massive stars are indeed the drivers of many important evolutionary processes in the Cosmos. At first sight this may be surprising since they are very rare! Let us recall that the initial mass function (IMF) is defined as the number of stars that has ever formed per unit area (disk) and per unit logarithmic mass interval (see e.g. Shapiro \& Teukolsky 1983). For instance a Salpeter IMF is given by
\begin{equation}
	\xi(\log M) d\log M=C {d M \over M^{2.35}}
\end{equation}
with $M$ in solar masses. Note that one can also write $\xi(\log M)=C M^{\Gamma}$ with $\Gamma=-1.35$.
Using this IMF, one can compute the number fraction of stars that ever formed in the Galaxy per unit area with an initial mass greater than $M$,
\begin{equation}
	N(M)={\int_{M}^{M_U}{d M \over M^{1-\Gamma}}\over\int_{M_L}^{M_U}{d M \over M^{1-\Gamma}}}={M_U^{\Gamma}-M^{\Gamma}\over M_U^{\Gamma}-M_L^{\Gamma}}.
\end{equation}
For $\Gamma=-1.35$, $M_U=120$ and $M_L=0.1$, the fraction number of stars with masses greater than 8 M$_\odot$
(usually taken as the mass superior limit for white dwarf formation) is 2.6 10$^{-3}$ (~3 stars over 1000).


In term of mass, the fraction of mass which is or was under the form of stars more massive than 8 M$_\odot$
is given by
\begin{equation}
	F(M)={\int_{M}^{M_U}{M d M \over M^{1-\Gamma}}\over\int_{M_L}^{M_U}{M d M \over M^{1-\Gamma}}}={M_U^{\Gamma+1}-M^{\Gamma+1}\over M_U^{\Gamma+1}-M_L^{\Gamma+1}}.
\end{equation}
With $\Gamma=-1.35$, this gives $\sim14$\%. Thus both in number and in mass, massive stars represent
small fractions.
What makes them nevertheless very important objects is that massive stars are very ``generous'' objects, injecting in the interstellar medium, in relatively short timescales (between ~3 and 30 million years) great
amounts of radiation, mass and mechanical energy:
\begin{itemize}
\item Radiation: using the mass-luminosity relation, a 100 M$_\odot$ has a luminosity about 1 million higher than the Sun. The high luminosity of massive stars allow them to be observed as individual objects well beyond the Local Group. Table~\ref{tab-1} gives the apparent magnitude in the visible of a B-type supergiant in different galaxies ordered by increasing distance. Knowing that the VLT magnitude limit is about 28.5, this means that
a B-type supergiant with an initial mass of about 25 M$_\odot$ can be viewed up to distances greater than 70 Mpc.

\begin{table}
	\centering
		\begin{tabular}{ccc}
\hline
  & & \\		
galaxy&	D[kpc]	& mv(B superG) \\
LMC	&  46	& 	8.5 \\
SMC	& 	63	& 	9.8 \\
M31	& 	724	& 	14.3 \\
M81	& 	3300	& 	17.7 \\
M100	& 	17000	& 	26.5 	\\
  & & \\
\hline  	
		\end{tabular}
	\caption{Apparent magnitude of a B-type supergiant in different galaxies.}
	\label{tab-1}
	\end{table}
	
Of course when the star explodes as a core collapse supernova, individual events can be seen at still much greater distance. Long soft GRB are believed to be associated to core collapse supernova events. The farthest event of that kind presently known is at a redshift of 6.26 (Cusumano et al.~2006)!

The collective effect of massive stars is of first importance to understand the light of galaxies. Typically about 2/3 of the visible light of galaxies arise from the massive star populations. 
Their high ionizing power give birth to HII regions which trace the regions of recent star formation. The strong UV luminosity of massive stars has also been (and is still) used to deduce the history of star formation in the Universe (see e.g. Hopkins \& Beacom 2006 for a discussion
of the cosmic star formation history). In some dusty galaxies, part of the UV light heats the dust and makes the galaxies to glow in IR (see e.g. the discussion of P\'erez-Gonz\'alez et al. 2006 on M81). Interestingly the ionizing front which expands with time around massive star can trigger star formation in their vicinity. Examples of such behavior is seen in the triffid nebula (see Hester et al. 2005).
The ionising flux of the Pop III stars played a key role in reionizing the early Universe
(Barkana 2006).
	
\item Mass: either through stellar winds or at the time of the supernova explosion great amounts of mass are injected back into the interstellar medium. As seen above, about 14\% of the mass of all the star ever formed,
called M$_*$ in the following, has been or is in the form of massive stars. Nearly all this mass has been, is, ejected back into the interstellar medium by massive stars (~13\% of M$_*$), only a very small amount (about 1\% of M$_*$) remains locked into compact remnants (neutron stars or black holes). If during the formation of a black hole, all the mass is swallowed by the black hole, smaller amounts of mass are returned. Typically, if all stars more massive than 30 M$_\odot$ follow such a scenario, then only a little more than 7\% of M$_*$ are returned. 
Part of the material returned, between 3.5 and 4.5\% of M$_*$, are under the form of new synthesized elements
and thus participe to the chemical evolution of the galaxies and of the universe as a whole.

\item Mechanical energy: during its lifetime a star with an initial mass of 60 M$_\odot$ will lose through stellar winds more than 3 fourths of its initial mass through stellar winds. Assuming a mass loss rate of
4 10$^{-5}$ M$_\odot$ per year, a velocity of the wind of 3000 km s$^{-1}$, one obtains a mechanical luminosity equal to 30000 solar luminosities or to about 10\% the radiation luminosity of the massive star. Integrating over the WR lifetime (about half a million years), one obtains that the mechanical energy injected into the surrounding amounts to 2 times 10$^{51}$ erg, {\it i.e.} a quantity of the same order of magnitude as the energy injected by a supernova explosion. Collective effects of stellar winds and supernova explosion may trigger in certain circumstances galactic superwinds (see the review by Veilleux et al. 2005). These galactic winds are also loaded in new chemical species and participate to the enrichment of the intergalactic medium or intra galactic medium of clusters of galaxies.
\end{itemize}
 
The points raised above underline the active role of massive stars in the Universe. As a consequence, in recent years, these objects have received and still receive a lot of attention from the part of observers.
In the following, we will present a few recent results bearing on the surface abundances, the mass loss by stellar winds, the rotational velocities, the surface magnetic fields, and the massive star populations in different environments.

\section{Surface abundances for probing stellar interiors}

Historically one of the first observational hint that indeed stars are building new elements came from the detection
of technetium at the surface of a red giant star (Merrill 1952). This radioactive element has indeed a disintegration lifetime  (213000 y) much shorter than the lifetime of the star in which it was observed. This clearly indicated that the origin of this element was processes occurring in the star itself. Today gamma ray line observations allow to detect gamma rays coming from the disintegration of radioactive elements. Up to now, five isotopes have been detected in this way: $^{26}$Al (mean lifetime 1.04 $\times$ 10$^6$ y) and $^{60}$Fe (2.2 $\times$ 10$^6$ y) have been detected as a diffuse emission in the galactic disk, their abundance reflect the global recent nucleosynthetic activity in the galactic disk in the last million years, while the emissions arising from the decay of $^{44}$Ti (89 y) $^{56}$Co (0.31 y) and $^{57}$Co (1.1 y) have been observed from point sources linked to young supernova remnants (Ti in Cas A, isotopes of Co in SN1987A) constraining the yield of individual events (see e.g. the review by Diehl et al. 2006).

The observation of surface abundances of non radioactive elements is also an efficient probe for studying stellar evolution and checking the internal nuclear processes. 
A star may present peculiar surface abundances (peculiar with respect to nearby otherwise similar stars) for many reasons:
\begin{itemize}
\item Internal processes may bring to the surface newly synthesized elements. This will constrain the internal mixing processes.
\item Mass loss by stellar winds or due to Roche Lobe overflow can uncover layers whose chemical composition has been modified by nuclear reactions.
\item The star may have accreted mass from an evolved companion, the companion having himself undergone change of its outer layers composition.
\item The star was formed from locally enriched material.
\end{itemize} 
Since the processes responsible for these changes of surface abundances may depend on the initial metal content, it is important to discuss observational evidences at different metallicities. We begin by discussing a few recent observational evidences at solar metallicity. 

\subsection{Evidences for mixing at solar metallicity}

Numerous observations show that mixing processes are active in radiative zones of massive stars. Of course it is not easy to discriminate between the various mechanisms outlined above and one can wonder whether some changes of the surface composition, attributed to mixing, cannot have another cause as mass loss or mass transfer in a close binary. There is at least one observation which seems to be very difficult to explain by another process than extra-mixing, this is the boron depletion at the surface of massive stars not accompanied by nitrogen enrichment (Venn et al. 2002; Mendel et al. 2006). Boron is a very fragile elements, destroyed by proton capture at temperatures ($< 6$ 10$^6$ K) well below those necessary to activate the CNO process. Such temperatures are reached already at about 1 M$_\odot$ below the surface in Main-Sequence B-type stars. Thus any process which would mix gently this upper layer would deplete boron while keeping the abundances of other less fragile elements, like for instance nitrogen, constant. Of course the mixing process responsible for the depletion of boron, if maintained on longer timescales, can be also responsible for nitrogen enhancement in more evolved stages. The above authors have observed stars (B-type stars in young clusters with initial masses between 8 and 14 M$_\odot$, for these stars mass loss during the MS phase is very weak), which are boron-depleted but not nitrogen rich.
These stars are at the beginning of their evolution having undergone only
a very mild mixing affecting only the upper layers. 
These authors have also observed
stars which are boron-depleted and nitrogen-rich.
These stars are in more advanced stages. 

The process of mass transfer cannot explain such a progressive mixing, since mass transfer in a close binary system would both result in depleting boron while simultaneously increasing the nitrogen abundance. 

Morel et al. (2006) have recently measured nitrogen enhancements at the surface of solar metallicity main-sequence B-type stars with initial masses between 9 and 20 M$_\odot$. Among the nine stars observed, two-thirds present N/C ratios which are between 1.5 and 3.7 the initial one (relative increase of ~0.2 - 0.6 dex). Such enrichment indicates that CNO processed material is mixed in the radiative envelope. Of course other processes as a mass transfer in a close binary could be invoked but there is no observational evidence that these stars belong to such systems. 

After the main sequence phase, it is more difficult to assign the changes of surface abundances to a mixing process operating in radiative zones. This comes from the fact that, at the red surpergiant phase, a deep outer convective envelope appears
which dredges up processed material at the surface. Any star having undergone
such dredge-up would present changes of its surface abundance, even if no extra-mixing
has been active in the previous phases. Of course stars crossing for the first time the HR-diagram from the blue to the red have not yet undergone such dredge-up and any changes of the surface abundances can be interpreted in this case as originating from a mixing process having occurred during the previous phases. However how to be sure that the star is on its first redwards track and
thus that the changes of the surface abundances are due only to an extra-mixing process?
At present there is no way to do this. May be in the future asteroseismology, by probing
the stellar interiors, will allow
to distinguish stars on their way to the red supergiant stage from stars having already undergone
a red supergiant stage. At the present time, we can only try to make comparisons with stellar models
and check if the surface abundances do reflect or not the effect of a red supergiant
dredge-up. 

\begin{figure}[!]
\includegraphics[width=2.5in,height=2.5in]{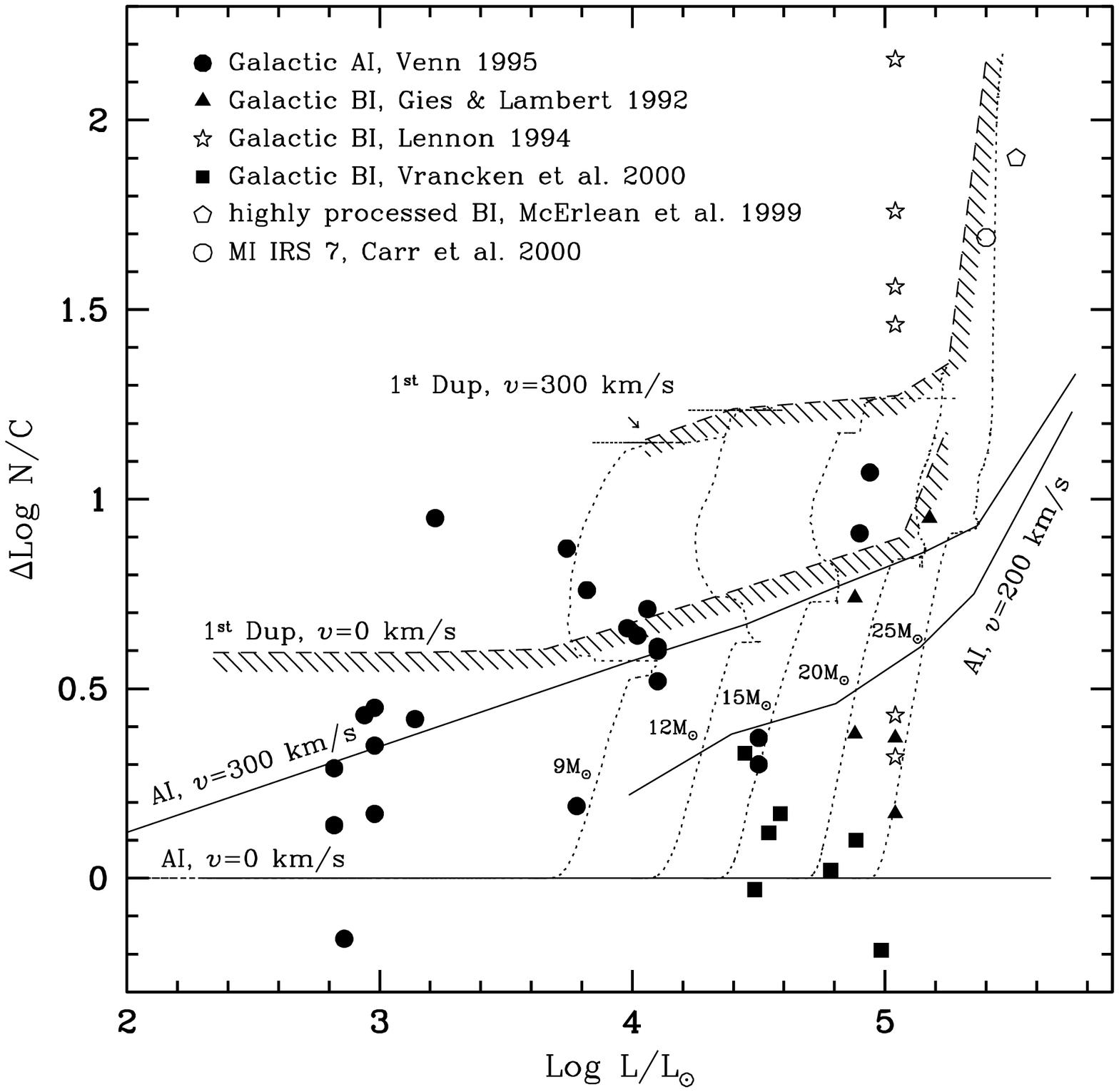}
\hfill
\includegraphics[width=2.5in,height=2.5in]{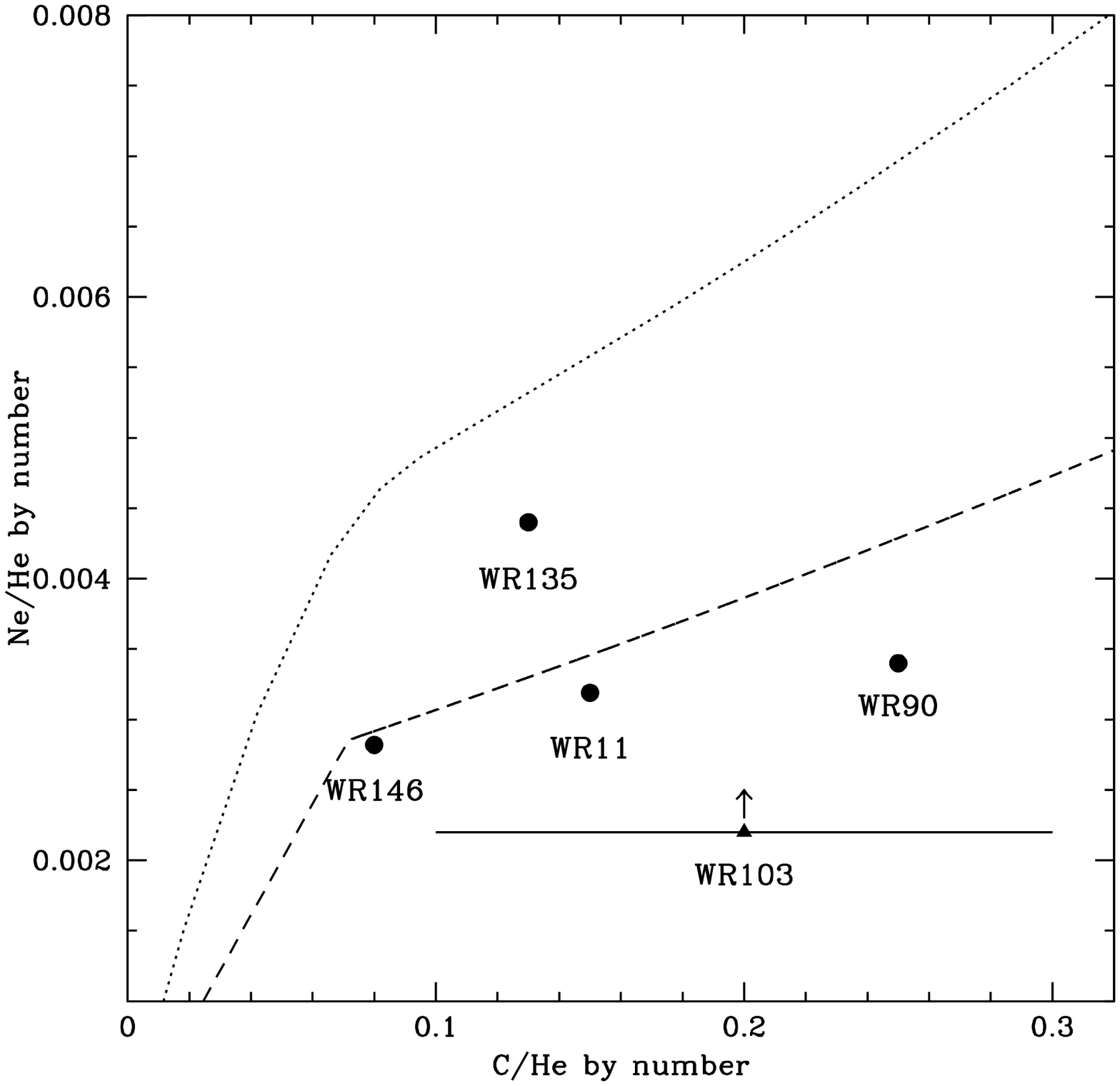}
\caption{{\it Left panel}:  Nitrogen to carbon ratios as a function of the luminosity. 
The quantity
$\Delta \log {\rm N\over \rm C}$ is equal to $\log ({\rm N/C})-\log ({\rm N/C})_i$, where N and C
are the surface abundances (in number) of nitrogen and carbon
respectively and the index $i$ indicates initial values.
The dotted lines represent evolutionary tracks for rotating models with initial
velocity of 300 km s$^{-1}$. 
The continuous lines indicate the $\Delta \log {\rm N\over \rm C}$ versus
$\log$~L/L$_\odot$ relation obtained from non-rotating and rotating models
when the star is an A-type supergiant (AI). The hatched lines show the same relation
but after the star has undergone the first dredge-up (1$^{st}$ Dup). The points 
represent the observed values (see text). 
{\it Right panel}: The black points show the Ne and C abundances observed at the surface
of WC stars by Dessart et al. (2000; filled circles) and by Crowther et al. (2006; filled triangle).
The dotted line show the prediction for a 60 M$_\odot$ stellar model with $Z$=0.020 (Meynet \& Maeder 2003) and the dashed line for a 60 M$_\odot$ stellar model with $Z$=0.014.}
\label{rsgSMC}
\end{figure}

The points in the left panel of Fig.~\ref{rsgSMC} correspond to observations of supergiants performed 
by Venn (1995), Gies and Lambert (1992), Lennon (1994), Vrancken et al. (2000),
McErlean et al. (1999) and Carr et al. (2000). 
We note that :

\noindent $\bullet$ The observed points are not concentrated along the horizontal line
$\Delta \log {\rm N/C}=0$ and along the line
labeled ``1$^{st}$ Dup, $\upsilon=$ 0 km/s''
as is predicted by standard (non--rotating) models.

\noindent $\bullet$ Some of the observed supergiants show nitrogen enhancements well below the values predicted by the first dredge--up of non--rotating models. These stars are probably,
as suggested by Venn (1995)
on their way from the MS to the red giant branch and have undergone some mixing
in the early stage of their evolution. 

\noindent $\bullet$ A-type supergiants observed by Venn (1995)
are also observed between the line corresponding to the first dredge--up of non--rotating models
and that of the first dredge--up of rotating models.
These stars might also be stars originating from relatively fast rotating stars evolving
towards the red supergiant stage.

\noindent $\bullet$ Very interestingly, B--type supergiants observed by
Lennon (1994) and McErlean et al. (1999)
present very high
N/C ratios.These
stars are likely on a blue loop after a red supergiant phase. Presently,
the models predict no blue loops at such high initial masses. Maybe this is an
indication that the mass loss rates during the red supergiant phase are much
higher than accounted for in the models, because high mass loss rates at this stage might bring the star back to the blue.

\noindent $\bullet$ We have indicated also the position in this diagram of the red
supergiant observed at the center of our Galaxy (IRS 7). This star shows a nitrogen enhancement well above what is predicted by standard models. According to Carr et al. (2000 and references therein) the metallicity at the galactic center is near solar. 
On the basis of this strong enrichment, these authors suggest that this star
might have been a rapid rotator. 

\noindent $\bullet$ Maximum values of about 1 dex are reached for luminosities
below about $\log L/L_\odot$ $\sim5$. For higher luminosities still higher enrichments are reached. 


Crowther et al (2006) have performed similar observations for blue supergiants ($\log T_{\rm eff}$ between 4.2 and 4.5) with initial masses between 20 and 40 M$_\odot$. They obtained also enrichments amounting to about 10 times the solar value for the N/C ratio and about 5 times the solar value for the N/O ratio. The greater enrichments appear to be realized for the stars with the lower effective temperatures (in the range indicated above).

\subsection{Evidences for mixing in the SMC}

Heap and Lanz (2003), Heap et al. (2005) have determined the nitrogen abundance at the surface of 18 O-type stars in the SMC (masses between 20 and 60 M$_\odot$). They found that 80\% of the stars present evidences for nitrogen enhancement, with typical N/C
enhancement ratios (relative to the SMC nebular value) ranging from 10 to 40.
Let us recall that the galactic sample of B-type stars studied by Morel et al. (2006) show nitrogen enhancements of at most a factor 4. Here the mass range is higher and this might indicate that the mixing process is more efficient in higher mass stars. It might also indicate that the process responsible for mixing is more efficient at low metallicity.
A feature which would support this last conclusion is the fact that the nitrogen enhancements appear to reach much higher values at the surface of SMC A-type supergiants than at the surface of similar stars in the Milky Way. Indeed Venn and Przybilla (2003) have shown that the maximum enhancement factor for the nitrogen to hydrogen ratios observed at the surface of SMC A-type supergiant is 40 (in number) while it is only 8 for Milky Way stars.
This is an indication that indeed mixing efficiency might be stronger at lower metallicities. As we shall see below, rotation might account for such a feature.

Mokiem et al. (2006) in the frame of the VLT-FLAMES survey have analysed the spectra of 31 O- and early B-type stars in the SMC, 21 of which are associated with the young massive cluster NGC346. The initial masses of the stars are comprised between 15 and about 100 M$_\odot$. They measured mass fractions of helium at the surface between 0.25 and  0.43 for Main Sequence stars and between
0.25 and 0.6 for supergiants. No standard evolutionary model can account for such strong He-enrichments.

\subsection{Changes of surface abundances due to mass loss and nuclear processing}

Wolf Rayet (WR) stars present emission lines in their spectra. These emission lines
are produced in an extended outflowing envelope (e.g. spectra of Wolf-Rayet
stars in the starburst galaxy IC10 can be seen in Crowther et al. 2003). 
Wolf--Rayet stars are nearly evaporating stars and thus are wonderful objects
to illustrate the effects of mass loss on massive star evolution.
Detailed reviews devoted to WR stars may be found in Abbott \& Conti
(1987),  Maeder \& Conti (1994), Willis (1999), Crowther (2007). 

Wolf--Rayet stars occupy the upper left corner of the HR diagram, they are hot and luminous
stars (see e.g. Hamann et al. 2006). They originate from stars more massive than about 30--40
M$_\odot$  that have lost their initial H--rich envelope  
by stellar winds or through a Roche lobe overflow in a close binary system.  
The stars enter the WR phase as WN stars, whose surface abundances are representative of
equilibrium CNO processed material. If the peeling off continues, the star may
enter the WC/WO phase, during which the He--burning products appear at the surface.
In our Galaxy, van der Hucht (2000; 2007) identifies 298  WR stars (171 WN, 10 WN/WC, 113 WC
and 4 WO), of which 24 (8\%) are in the open cluster Westerlund 1, and 60 (20\%) are in open clusters near the
Galactic Center. One estimates that
their total number in our Galaxy is as high as a few thousands.

The observed surface abundances during the WN phase correspond to CNO equilibrium values, while those
observed during the WC/WO phase well correspond to the apparition at the surface of He-burning products (see the review by Crowther 2007).
In particular the high overabundance of $^{22}$Ne at the surface of the
WC star predicted by He-burning reactions is well confirmed by the observations (Willis 1999; Dessart et al. 2000; Crowther 2006). Note that the abundance of $^{22}$Ne at this stage (WC) is an indication
of the initial CNO content of the star. Indeed, $^{22}$Ne comes mainly from the destruction of $^{14}$N at the beginning of the
helium burning phase, this $^{14}$N being the result of the transformation of carbon and oxygen into nitrogen operated by the CNO cycle during the core H-burning phase.
In that respect it is interesting to note that comparison between observed Ne/He ratio at the surface of WC star with models computed with Z=0.02 show that models overpredict the Ne abundance, while models starting
with a lower initial metallicity gives a much better fit as can be seen in the right panel of Fig.~\ref{rsgSMC}. This tends to
confirm that massive star in the solar neighborhood have initial metallicities in agreement with the
Asplund et al. (2005) solar abundances.
Let us note that this overabundance of $^{22}$Ne at the surface of WC stars is not only an important confirmation of the nuclear reaction chains occurring during He-burning, but is also related to
the question of the origin of the material accelerated into galactic cosmic rays (see recent
measurements of the $^{22}$Ne/$^{20}$Ne ratio in cosmic rays in Binns et al. 2005) and to the weak
s process in massive stars since $^{22}$Ne is the source of neutrons in these stars  (see the lecture by Arnould this volume).

Thus a general agreement is found between observations and the theoretical predictions.
However some WR stars (about 4.4\% of the galactic WR stars, van der Hucht 2001) show simultaneously 
enhancements of H- and He-burning products.
These stars are not explained
by standard models. In addition to mass loss, some mixing process has to be invoked for explaining
these stars (see discussion later in this paper). 

\subsection{Mass accretion from an evolved companion}

Can a massive star obtain peculiar surface abundances due to the fact that it belongs to a close binary system? This in principle could be achieved in three ways: first tidal mixing might induce internal mixing and provokes changes of the surface abundance even before any mass transfer, second the primary may lose through Roche lobe overflow part or all of its envelope and thus make its internal core uncovered, third the secondary may receive from the primary some material whose composition has been changed during the evolution of the primary. 

Very little is known about tidal mixing and its efficiency. This is probably an area of research where still a lot has to be done both from the point of view of the observations and from the point of view of theory. 

Concerning the second point (mass loss by Roche lobe overflow), we have already mentioned above that some WR stars might be formed through this channel. The question is how many WR stars owe their WR characteristics to mass transfer? 
At low metallicity, it has generally been thought that WR stars might preferentially be formed by mass transfer
through Roche Lobe Overflow in close binary systems. 
The main reason is that, at low metallicity, the mass loss rates are much lower than at higher metallicity, 
making thus more difficult the ejection of the H--rich envelope by stellar winds.
However this idea has been challenged by the works of Foellmi et al. (2003ab). They
looked for periodic radial velocity variability from all the WR stars in the Small Magellanic Cloud and from two thirds of the WR stars in
the Large Magellanic Cloud. They found that the percentage of binaries among the WR stars is of the order of 40\%
for the SMC and 30\% for the LMC. These percentages are comparable or even below the percentage of binaries among the 
WR stars in our Galaxy. It means that, in the SMC, at most 40\% of the WR stars could originate from
mass transfer through Roche Lobe Overflow (RLOF) in a close binary system. The real fraction is smaller since
RLOF has not necessarily occurred in all these systems.
Indeed some fractions of the WR stars in binaries belong to systems where the two components are sufficiently far
away from each other that binarity does not affect their evolution.
Therefore even in the SMC, the majority of the WR stars originates via the single star scenario.

Finally can a massive star accrete material from an evolved companion and thus acquire a peculiar surface abundance? Probably this occurs in nature, but in systems where both components are massive stars, i.e. stars with strong stellar winds, accretion onto the secondary might be prevented by the pressure exerted by the wind of the secondary.
A spectacular example of a colliding wind system is the WR104 binary system, consisting of a WC and an OB companion. A dust formation zone is associated with the region of the collision between the stellar winds.
The images in infrared taken by  
Tuthill et al. (1999) with the Keck I telescope show that the dust forms a spiral trajectory with a rotation period of 220$\pm$30 days. WR104 undergoes strong dust formation episodes when the companion star goes through periastron in a highly elliptical orbit.
Dust formation is possible thanks to the wind compression associated with the bow shock between the stellar winds. 
Interestingly similar systems have been found in the Quintuplet cluster in the galactic center (Tuthill et al. 2006).

Let us end this section by mentioning an interesting approach to constrain the chemical composition of supernova ejecta. When a core collapse occurs in a binary system, there is some chance that part of the ejecta be intercepted by the companion. In that case the companion may present surface abundances different from that of nearby stars. At the present time  three such cases have been observed (Israelian et al. 1999; Gonz\'alez Hern\'andez et al. 2004, 2005). For instance large overabundances (factors between 6 to 10 with respect to solar abundances) of oxygen, magnesium, silicon and sulphur have been observed in the atmosphere of the star orbiting a probable black hole Nova Scorpii 1994 (Israelian et al. 1999). This would be the first observational evidence that a
supernova event is associated to the birth of a black hole. This would also indicate that not all the mass is swallowed by the black hole, part of it participate to the chemical enrichment of the
surrounding.

\subsection{Primordial peculiar abundance patterns}

Some chemical characteristics of stars are inherited from the protostellar cloud from which they formed. For instance in most of the cases the stellar content in iron is inherited from the iron content of the cloud from which the star formed. Now there are some evidences that some chemical anomalies involving lighter elements (anomaly here in the sense that the star presents a different abundance pattern than otherwise similar stars) might also be inherited from the protostellar cloud composition. For instance, the high abundances in carbon and oxygen shown by about 20\% of the very metal poor halo stars ([Fe/H]$< -2.0$, see e.g. Beers \& Christlieb 2005; Lucatello et al. 2006), or the sodium-rich and oxygen depleted stars in globular clusters (see e.g. Carreta et al. 2006). In that case, at least part of the peculiar abundance pattern does not result from internal processes occurring in the observed star, but result from localized injection of matter (localized otherwise there would be identity of the composition for all the stars) processed by a previous generation of stars.

\begin{figure}[tb]
\begin{center}
  \resizebox{10cm}{!}{\includegraphics[angle=-90]{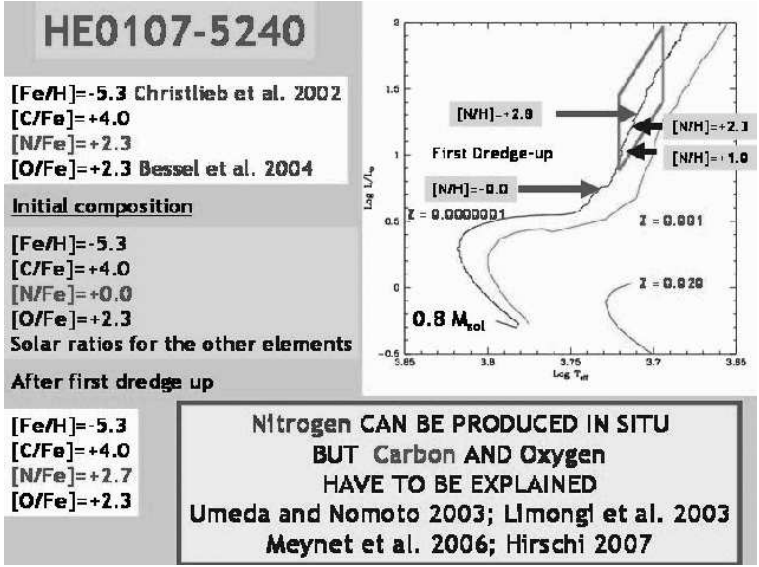}}
  \caption{Evolutionary tracks of a 0.8 M$_\odot$ with initially enhanced carbon and oxygen abundance and for
  Z=10$^{-7}$. Partial tracks at two other metallicities are shown for comparison. 
}
  \label{Cristlieb}
\end{center}  
\end{figure}

The abundances of some elements in a C-rich metal poor halo star (Christlieb et al. 2002; see also Bessel et al. 2004) is given
in Fig.~\ref{Cristlieb}. The star is a red giant, thus one cannot exclude that part of the surface abundances result from internal mixing. For instance, as shown in Fig.~\ref{Cristlieb}, the very high nitrogen surface abundance might be due to the effect of the first dredge-up provided the star began its evolution with high carbon and oxygen abundance. Of course the nitrogen overabundance might also be inherited in part from a high initial nitrogen abundance. This is at least the case for some of the C-rich stars as for instance
the dwarf
star G77-61 (Plez \& Cohen 2005).
The initial mass of G77-61 is estimated to be between 0.3 and 0.5 M$_\odot$, it has an [Fe/H]=-4.03,
[C/Fe]=2.6, [N/Fe]=2.6, and a $^{12}$C/$^{13}$C ratio of 5$\pm$1.
In this case, there is no way for the star, which burns
its hydrogen through the pp chains to produce nitrogen. 
There is even less possibility to produce surface enhancements of carbon and oxygen. 
Therefore, the ``in situ'' scenario can be excluded at least for this star.
In that case, the observed abundance must have been inherited either from the protostellar cloud or from a
companion through a mass transfer episode. 

\section{Mass loss rates by massive stars}

\subsection{Some generalities about stellar winds}

The intensity of the mass loss is a key quantity for stellar evolution. It affects the tracks in the HR diagram, lifetimes, surface
abundances, chemical yields, presupernova status, the nature of the remnant,
the mechanical energy released in the interstellar medium, the hardness of the radiation field, etc...  

The main trigger of mass loss is radiation pressure.
As written by Eddington (1926) `` {\it … the radiation observed to be emitted must work
its way through the star, and if there were too much
obstruction it would blow up the star.}'' 
Note that if it was realized already in the 20s that radiation pressure may produce mass loss, it is only when, in the late 60s, sensitive UV diagnostics of mass loss from O-star were available that the effects of mass loss on the evolution of stars were really considered. 

Radiation triggers mass loss through the line opacities in hot stars. 
It may also power strong mass loss through the continuum opacity when the star is near the Eddington limit. For cool stars, radiation pressure is exerted on the dust.

For hot stars, typical values for the terminal wind velocity, $\upsilon_\infty$ is of the order
of 3 times the escape velocity, {\it i.e.} about 2000-3000 km/s,  mass loss rates are between 10$^{-8}$-10$^{-4}$ M$_\odot$ per year. Luminous Blue Variable (LBV) stars show during outbursts mass loss rates as high as 10$^{-4}$-10$^{-1}$ M$_\odot$ per year.

According to the  recent mass loss rates \mdot\ behave with luminosity $L$
like 

\begin{eqnarray}
\mdot\  \; \sim \; L^{1.7} \; ,
\end{eqnarray}

\noindent
with the mass--luminosity relation for massive stars, this gives

\begin{eqnarray}
L \; \sim \; M^2 \quad \rightarrow \quad \mdot\ \; \sim \; M^{3.4} \;.
\end{eqnarray} 

\noindent
From this we may  estimate the typical timescale for mass loss

\begin{eqnarray}
t_{\mdot} \; \sim \; \frac{M}{\mdot } \; \sim \; \frac{1}{M^{2.4}} \;.
\label{tmdot}
\end{eqnarray}

\noindent
This is to be compared to the MS lifetime
$t_{\mathrm{MS}}$. For massive stars, it scales like 

\begin{eqnarray}
t_{\mathrm{MS}} \; \sim \; M^{-0.6}   \; ,
\label{tms}
\end{eqnarray}

\noindent
which shows that with increasing mass the timescale for mass loss decreases
much faster than the MS lifetime. One can also estimate the behavior of the
 amount $\Delta M$ of mass lost  with stellar mass 

\begin{eqnarray}
\Delta M \; \sim \; M^{2.8} \quad \rightarrow \quad \frac{\Delta M}{M} \; \sim \; M^{1.8} \; .
\end{eqnarray}

\noindent
Thus, not only the amount of mass lost grows with the stellar mass, but even
the relative amount of mass loss grows fast with stellar masses, which illustrates the importance of this effect.

In addition to the intensity of the stellar winds for different evolutionary phases, one needs to know 
how the winds vary with the metallicity. This is a key effect to understand the different massive star populations observed in regions of different metallicities. This has also an important impact on the yields expected from stellar models at various metallicities.

Current wisdom considers that very metal poor stars lose no or very small amounts
of mass through radiatively driven stellar winds. This comes from the fact that when the metallicity
is low, the number of absorbing lines is small and thus the coupling between the radiative forces and the matter is weak. Wind models imposes a scaling relation of the
type 
\begin{equation}
\dot M(Z)=\left({Z \over Z_\odot} \right)^\alpha\dot M(Z_\odot),
\end{equation}
where $\dot M(Z)$ is the mass loss rate when the metallicitity is equal to $Z$ and $\dot M(Z_\odot)$
is the mass loss rate for the solar metallicity, $Z$ being the mass fraction of heavy elements.
In the metallicity range from 1/30 to 3.0 times solar,
the value of $\alpha$ is between 0.5 and 0.8 according to stellar wind models (Kudritzki et al. 1987; Leitherer et al. 1992; Vink et al. 2001). 
Such a scaling law implies for instance that 
a non-rotating 60 M$_\odot$ with $Z=0.02$
ends its stellar life with a final mass of 14.6 M$_\odot$, the same model with a metallicity of
$Z=0.00001$ ends its lifetime with a mass of 59.57 M$_\odot$ (cf. models of Meynet \& Meader 2005 and Meynet et al. 2006 with $\alpha=0.5$). 

Thus
one can expect that the metal poor 60 M$_\odot$ star will give birth to a black hole.
In that case nearly all (if not all) the stellar mass may disappear in the remnant preventing the star from enriching the
interstellar medium in new synthesized elements (see however above the case of Nova Scorpii). The 
metal rich model on the other hand will probably leave a neutron star and contribute to the enrichment of the ISM through both
the wind and supernova ejecta. 

A star which loses a lot of material by stellar winds may differently enrich the interstellar medium
in new elements, compared to a star which would have retained its mass all along until the supernova explosion. As Maeder (1992) pointed out, when the stellar winds are strong, material partially processed
by the nuclear reactions will be released, favoring some species (as helium and carbon which would be otherwise partially destroyed if
remained locked into the star) and disfavoring other ones (as e.g. oxygen which would be produced by further transformation of the species which are wind-ejected).

\subsection{Some recent mass loss determinations}

There are mainly three types of observations which can be used for determining mass loss rates:
\begin{enumerate}
\item Free-free continuum emission at radio wavelengths.
\item H$\alpha$ line emission.
\item UV resonant-line absorption (P-cygni profiles).
\end{enumerate}
These three methods probe different portion of the stellar winds: radio emission probes the very distant rarefied, constant velocity regions, H$\alpha$ emission probes the near-star rapidly accelerating region,
while the UV probes nearly the whole zone in between. The first two diagnostics are emission processes which involve two particles. Thus these emission processes vary as the square of the density of the emitting region.
The last process depends on the density. This is important to realize since the first two
diagnostics do not show the same sensitivity to the clumping effect than the last one.

Let us suppose that  the matter in the wind is not distributed smoothly but aggregates into clumps of high density. What will be the effect of this clumping if one uses smooth stellar wind models to determine the mass loss rates? Mass loss rates are determined by determining the wind model which correctly reproduces the intensity level of the emission either in the radio regime or of the H$\alpha$ emission line.
Using a smooth stellar winds, the excess emissions arising from the inhomogeneities will be uncorrectly interpreted as arising from a smooth but denser medium. This will overestimate the mass loss rates.
If $f$ is the filling factor of the denser component, comprised between 0 and 1, and $x$ the density ratio between the interclump medium and the clumps, Abbott et al. (1981) showed that the mass loss rates obtained from $\rho^2$-diagnostics, supposing (wrongly) a smooth wind, are a factor $\left( { f+(1-f)x^2\over [f+x(1-f)]^2} \right)^{1/2}$ greater than the actual mass loss of a clumpy wind. If $x=0$ (no matter between the clumps), then
the enhancement factor is equal to $1/\sqrt{f}$.

The UV resonant-line absorption diagnostic are much less sensitive to clumping. Indeed absorption depends on the integrated quantity of matter along the line of sight. As long as the clumps are optically thin, this integral is not sensitive to the mass distribution and therefore this kind of diagnostic is much less sensitive to clumping (Note that the ionisation fraction of the element responsible for the  line is affected by clumps in the wind thus this diagnostic is not completely free of clumping effect).

Comparing mass loss rates obtained by different technics, sometimes very important differences have  been obtained.
For instance, Fullerton et al. (2006) using UV line of P$^{+4}$ obtained mass loss rates reduced by  a factor ten or more with respect to mass loss determination from radio or H$\alpha$ determination. Bouret et al. (2005)
obtained qualitatively similar results than Fullerton et al. (2006) but with considerable lower reduction factor (about 3). Such reduction of the mass loss rates during the O-type star phase may have important consequences. Typically a 120 M$_\odot$ loses during its 2.5 Myr about 50 M$_\odot$ when mass loss rates of the order of 2 10$^{-5}$ M$_\odot$ per year
are used. Dividing this mass loss rate by 10, would imply that in the same period, the star would lose only 5 M$_\odot$! Unless stars are strongly mixed (by e.g. fast rotation), or that all WR stars originate in binary systems (but see above), it would be difficult to understand how WR stars form.

Such low mass loss rates need to be confirmed by further studies. One points of concern
is that in many models accounting for clumping it is assumed that there is no mass between
the clumps, which is probably incorrect and lead to an underestimate of the mass loss. In any case, this is presumably too early at the present time to consider these very low mass loss rates as the canonical mass loss rates to be used in stellar models. It is however worthwhile to keep in mind this problem and to keep tuned on what is going on in the next few years (see the interesting discussion in Puls et al. 2006). 

As already mentioned above, LBV stars, during outbursts, show extremely high mass loss rates. For instance $\eta$ Carinae ejected near the middle of the eighteenth century
between 12 and 20 M$_\odot$ in a period of 20 years, giving an average mass loss rate during this period of 0.5 M$_\odot$ per year.
Such a high mass loss cannot be only radiatively driven according to Owocki et al. (2004). These authors have shown that the maximum mass loss rate that radiation can driven is given by
$$\dot M\sim 1.4 \times 10^{-4} L_6 {\rm M}_\odot {\rm yr}^{-1},$$
with $L_6$ the luminosity expressed in unit of 10$^6$ L$_\odot$. 
This means that for $L_6=5$ (about the case of $\eta$Car) the maximum mass loss rate would be less than 10$^{-3}$ M$_\odot$ per year, well below the mass loss during the outbursts. These outbursts, which are more shell ejections than steady stellar winds, involve other processes in addition to the effects of the radiation pressure. Among the models proposed let us mention the geyser model by Maeder (1992a), or the reaching of the $\Omega\Gamma$-limit (Maeder \& Meynet 2000).

Smith \& Owocki (2006) suggest that the very massive stars might lose most of their mass during such LBV outbursts, {i.e.} might have their winds ``dominated by optically thick, continuum- driven outbursts or explosions, instead of by steady line-driven winds''. While it is not
possible to discard at the present time such a possibility, certainly such a strong conclusion should await confirmation of the very low mass loss rates discussed above. Presently we are still far from having reached a consensus on that point. Independently of that, we can wonder to which
extent the LBV outburst depends on the metallicity. If the mechanism is mainly triggered by 
continuum opacity, we can expect that there is only a weak or may be no dependence on the metallicity. This may allow Pop III stars to lose some significant amount of material through this mechanism.

After the LBV phase, massive stars evolve into the Wolf-Rayet phase, also characterized by strong mass loss rates. Many recent grids of stellar models use the recipe given by Nugis and Lamers (2000) for the WR mass loss rates. These authors deduced the mass loss rates from radio emission power and accounted for the clumping effects. Until very recently, it was considered that the WR mass loss rates were independent on the initial metallicity {i.e.} that a WN stars in the SMC, LMC and
in the galaxy would lose mass at the same rate provided they have the same luminosity and the same surface abundances. This view is now challenged. Vink \& de Koter (2005) find that the winds of WN stars are mainly triggered by iron lines. They suggest a dependence of mass loss on $Z$ (initial value) similar to that of massive OB stars. According to these authors, the winds of WC stars is also found to be dependent on the iron abundance, but in this case, the metallicity dependence is less steep than for OB stars.
Their results apply over a range of metallicities given by 10$^{-5} \le (Z/Z_\odot) \le 10$. 
Very interestingly, they find that once the metal abundance drops below $(Z/Z_\odot) \sim 10^{-3}$,
mass loss of WC stars no longer declines. This is due to an increased importance of radiative driving by intermediate mass elements, such as carbon. These results have profound consequences for the evolution of stars at low metallicity, affecting the predicted Wolf-Rayet populations
(Eldridge \& Vink 2006), the evolution of the progenitors of collapsars and long soft Gamma Ray Bursts (Yoon \& Langer 2005; Woosley \& Heger 2006; Meynet \& Maeder 2007).

Lower initial mass stars evolve to the red supergiant stage where mass loss is enhanced with respect to the mass loss rates in the blue part of the HR diagram (see for instance de Jager et al. 1988). In this evolutionary stage, determination of the mass loss is still more difficult than in the blue part of the HR diagram due in part to the presence of dust and to
various instabilities active in red supergiant atmospheres (e.g. convection becomes supersonic and
turbulent pressure can no long be ignored). An illustration of the difficulty comes from the recent determination of red
supergiant mass loss rates by van Loon et al. (2005). Their study is based on the analysis of optical spectra of a sample of dust-enshrouded red giants in the LMC, complemented with spectroscopic and infrared photometric data from the literature. Comparison with galactic
AGB stars and red supergiants show excellent agreement for dust-enshrouded objects, but not
for optically bright ones. This indicates that their recipe only applies to dust-enshrouded stars.
If applied to objects which are not dust enshrouded, their formula gives 
values which are overestimated by a factor 3-50! In this context the questions of which stars do
become dust-enshrouded, at which stage, for how long, become critical to make correct prediction on the mass lost by stellar winds. One can also note a very interesting point deduced from the study
by van Loon et al. (2005), that for dust-enshrouded objects mass loss appears to be independent of the metallicity! This may also have very important consequences for our understanding of metal poor red supergiant stars.

\section{Rotational velocities}

Observation of the rotation of stars is a very old subject. The first known detailed account of the rotation of the Sun dates back to the time of Galileo Galilei at the beginning of the seventeenth century. Nowadays our knowledge of the rotation of stars is mainly based on doppler
widening of absorption lines. Very recently, interferometric technics allowed to observe the
shape of some stars strongly deformed by fast rotation as Archernar (Domiciano de Souza et al. 2003)
or to measure the variation with the latitude of the effective temperature 
(Ohishi et al. 2004; Domiciano de Souza et al. 2005) in order to test the
von Zeipel theorem (von Zeipel 1924). 

The measure of the rotational velocity through the Doppler effect is of course not free from spurious effects. First this technic gives access only to $\upsilon \sin i$ where $i$ is the angle between the line of sight and the rotational axis. This quantity
cannot be known except if some other observations as e.g. a rotational period obtained from variability 
induced by the presence of dark spot at the surface are available. But even in that case, the knowledge of $\upsilon$ require the knowledge of the radius (more precisely of the equatorial radius), moreover if the surface rotates differentially like the solar surface, the period thus obtained might only be an average over the surface. Second due to the von Zeipel theorem, the fast rotating regions of the star, i.e. the zones near the equator, are less luminous than the polar regions which have small linear velocities. This effect produces an underestimate of the rotational velocities. Townsend et al. (2004) obtain the following behavior between the equivalent width of the HeI4471 
line for a B2-type star and the equatorial
velocity: the relation is nearly perfectly linear for $\sin i=1$ and $\upsilon/\upsilon_{\rm crit}$ inferior to 70\% (for a solar metallicity 9 M$_\odot$
star, this corresponds to velocities inferior to 300 km s$^{-1}$ on the ZAMS). 
In that velocity regime, the equivalent width measurements, due to the above effects,
underestimate the velocities by at most 3-4\%. For  $\upsilon/\upsilon_{\rm crit} > 70$\%, the relation is no long linear, the equivalent width increases much more slowly with the velocity, it even goes through a maximum, decreasing a little for velocities near the critical limit. This means that in that case
the equivalent width becomes a poor indicator of the surface velocity, a star rotating
at the critical limit presenting the same line broadening as a star rotating at 80\% the critical velocity! Said in other words, this technic may underestimate the real velocity of the star by 20-25\% for very fast rotators (typically for $\upsilon/\upsilon_{\rm crit} > 75$\%, {\it i.e.}
for velocities superior to about 340 km s$^{-1}$, for a 9M$_\odot$ at solar metallicity).
This effect might be the reason why the rotational velocity of Be stars that are considered by many
authors as stars rotating at the critical limit, have measured velocities corresponding to only
about 70\% of the critical limit.

Recent works have performed new estimates of the rotation rate of B-type stars. Dufton et al. (2006)
have measured the rotation of stars in two galactic clusters NGC 3293 and NGC 4755 whose ages are
comprised between 10 and 15 My. For stars with masses between 3 and 12 M$_\odot$, they obtain that
the distribution of velocities peak at 250 km s$^{-1}$ accounting for a random distribution of the inclination\footnote{In that case, one passes from the $\upsilon\sin i$ values to the
$\upsilon$ values by multiplying  the $\upsilon\sin i$ by $4/\pi=1.27$. This is true only on average.}. Similar results have been obtained by 
Huang \& Gies (2006). These last authors noticed that there is a deficit of slow rotators in clusters
with respect to field stars. They show also that the average velocity of stars with masses between
8.5 and 16 M$_\odot$ remains more or less constant when the surface polar gravity decreases. i.e. when evolution proceeds, while for masses between 2.5 and 8.5 M$_\odot$ the velocity decreases with
decreasing polar gravity, indicating that in this mass range, some breaking mechanism is active which does not work in the higher mass range (magnetic breaking?). Binary systems appear to experience more spin down than single stars. 

Huang \& Gies (2006) have also observed that for the most massive stars of their sample (masses
between 8.5 and 16 M$_\odot$) there is an increase of the mass fraction of helium at the surface of the star when the polar gravity decreases. The increase of $Y$, the mass fraction of helium, amounts to 23$\pm$13\%, between the ZAMS and the TAMS (i.e. if $Y$=0.25 on the ZAMS, at the TAMS, $Y$ may 
reach values as high as 0.31). Such helium enrichment at the surface have also been obtained
by Lyubimkov et al. (2004).

Does the distribution of the velocities vary as a function of the metallicity?
This question is still debated. For instance Keller (2004) measured the rotational velocity of
100 MS early B-type stars in clusters of the LMC and compared the results with observations of early B-type stars in clusters of the solar neighborhood. He obtains that the LMC stars are faster rotators
than the galactic ones: the mean values of $\upsilon\sin i$ is 116 km s$^{-1}$ for the galactic stars
and 146 km s$^{-1}$ for the LMC stars. On the other hand, Penny et al. (2004) find no difference between the velocities of O-type stars in the Magellanic Clouds and in the Galaxy. Numerous surveys
are now being undertaken with aim to provide further constraints on this topic, as e.g. the VLT-Flames survey. At this stage, we can just mention that the fact that the fraction of Be stars
increases with the metallicity (Be stars being stars near the break-up limit, see Sect. 7.1) may favor the
situation where the distribution of velocity contains more fast rotators at low metallicity.


\section{Magnetic fields}

A review of the observations of magnetic fields of OB stars can be found in Henrichs et al. (2005).
Many indirect observations, like wind variabilities, chemical peculiarity, specific pulsation behavior, anomalous X-ray emission, non-thermal emission in the radio region indicate that OB stars have a surface magnetic field. Only for a few of them (see table~\ref{OBH}) was it possible to obtain a direct magnetic detection. This however cannot be interpreted as an absence of magnetic field for those for which no magnetic field was detected. These measures are very delicate. The accuracy in the magnetic field for a star with a given magnitude varies with the square root of the number of lines available and with the inverse of the square root of the rotational broadening
of the line. Thus the measures become more and more difficult when earlier spectral type are considered (smaller number of lines and fast rotation).

\begin{table}[t]
\caption{Measured magnetic fields of OB stars. Note that only two magnetic O-type stars are known.
The angle between the magnetic and rotation axis is $\beta$.} \label{tbl-1}
\begin{center}\scriptsize
\begin{tabular}{lclrrrrrr}
\hline
 & & & & & & & & \\
Star    &   Ref.  & Sp. ty.& $\upsilon\sin i$ & P$_{\rm rot}$ & Mass       & Incl. & b   & B$_{\rm pol}$ \\
        &         &        &  km s$^{-1}$     &  days         & M$_\odot$  & deg.  & deg.& G.    \\
 & & & & & & & & \\
HD191612       & (6) &       &       & 538    &     &           &  45        & ~1500        \\
$\Theta$ Ori C & (1) & O4-6V & 20    & 15.4   &  45 & 45        &  42$\pm$6  & 1100$\pm$100 \\
$\beta$ Cep    & (2) & B1IVe & 27    & 12     &  12 & 60$\pm$10 &  85$\pm$10 & 360$\pm$40    \\
$\tau$ Sco     & (7) & B0.2V &       & 41     &     &           &            & ~500          \\
V2052 Oph      & (3) & B1V   & 63    & 3.64   &  10 & 71$\pm$10 &  35$\pm$17 & 250$\pm$190    \\
$\xi$ Cas     & (4) & B2IV  & 17    & 5.37   &   9 & 18$\pm$4  &  80$\pm$4  & 340$\pm$90     \\
$\omega$ Ori   & (5) & B2IVe &172    & 1.29   &   8 & 42$\pm$7  &  50$\pm$25 & 530$\pm$200    \\
He-peculiar    &     & B1-B8p&       & 0.9-22 & <10 &           &            & 1000-10000     \\
 & & & & & & & & \\         
\hline
 & & & & & & & & \\  
\multispan9{(1) Donati et al. 2002 (2) Henrichs et al. 2000 (3,4,5) Neiner et al. 2003abc,\hfill}   \\
\multispan9{(6,7) Donati et al. 2006ab\hfill}
\end{tabular}
\label{OBH}
\end{center}
\end{table}

From the values given in Table~\ref{OBH}, one can make two remarks based on order of magnitude
arguments. First the magnetic field of pulsars is of the order of 10$^{12}$ G (see. e.g. Srinivasan 1995). In case the magnetic flux remains constant during the stellar lifetimes, such a magnetic
field implies a magnetic field of the order of about 10 G during the MS phase [(10 km/$R)^2 \times 10^{12}$, where $R$ is the radius of the star on the MS phase, here taken equal to 5$R_\odot$]. Thus,
the measured magnetic fluxes are one to two orders of magnitude greater than those required to reproduce the magnetic fields of pulsars. The observed values are more compatible with magnetars (see e.g. Gotz et al. 2007). A second point interesting to underline is that
such high values of the magnetic field probably strongly affects the stellar winds. Ud-Doula \& Owocki (2002) have suggested that in case $\eta=(B^2/8\pi)/(\rho v^2/2) > 1$, i.e. if the energy density in the field is superior to the kinetic energy density in the wind ($v$ is here the wind velocity), then
the magnetic field has an impact on the winds. From this formula, one obtains that for early-type stars, for a field $B\sim 50-100$ G, one has $\eta > 1$. One sees that this is clearly the case for the stars
listed in table~\ref{OBH}, however, as discussed above, these stars might not be representative of the bulk of OB stars. 

All magnetic B stars appear to have some abundance anomaly (see table 2 in Henrichs et al. 2005):
typically V2052 Oph has a N/C value of 1.9 $\times$ (N/C)$_\odot$, $\xi$Cas and $\omega$Ori of
2.9, 1.8 $\times$ (N/C)$_\odot$ (number ratios, the solar values are those of Grevesse \& Sauval 1998).

\section{Populations of massive stars in galaxies}

Theoretical stellar models allow to find evolutionary filiations between different types of massive stars. The different scenarios are indicated below. 
\vskip 5mm
\noindent
{\bf \underline{{$M >90  M_{\odot}$}}}:  O –- Of –- WNL –- (WNE) -– WCL –- WCE -– 
SN (Hypernova low Z  ?)\\
\noindent
{\bf \underline{{$60-90 \; M_{\odot}$}}}: O –- Of/WNL$\Leftrightarrow$LBV -– WNL(H poor)-– WCL-E -– SN(SNIIn?)\\
\noindent
{\bf \underline{{$40-60 \; M_{\odot}$}}}: O –- BSG –-  LBV $\Leftrightarrow$ WNL -–(WNE) -- WCL-E –- SN(SNIb) \\
\hspace*{5.9cm}  - WCL-E - WO – SN (SNIc) \\
\noindent
{\bf \underline{{$30-40 \; M_{\odot}$}}}:  O –- BSG –- RSG  --  WNE –- WCE -– SN(SNIb)\\
\hspace*{4.0cm}                        OH/IR $\Leftrightarrow$ LBV ? \\
\noindent
{\bf \underline{{$25-30 \; M_{\odot}$}}}: O -–(BSG)–-  RSG  -- BSG (blue loop) -- RSG  -- SN(SNIIb, SNIIL)\\
\noindent
{\bf \underline{{$10-25 \; M_{\odot}$}}}: O –-  RSG -– (Cepheid loop, $M < 15 \; M_{\odot}$) – RSG -- 
SN (SNIIL, SNIIp)\\ 

The sign $\Leftrightarrow$ means back and forth motions between the two  stages. The various types of supernovae are tentatively indicated. The limits 
between the various scenarios  depend on metallicity $Z$ and rotation. 

For constraining stellar models, the best way is to compare the results of tailored theoretical models with well observed characteristics of single stars. The case of the Sun is in that
respect exemplar. Considering observed populations of massive stars for constraining stellar models may appear at first sight a loose way to proceed since in addition to the physics of stars, other
ingredients enter into the comparison as the Initial Mass Function (IMF) and the Star Formation History.
However in some circumstances, observed stellar populations may provide powerful
constraints. For instance, IMF has no influence on ratios of massive stars 
involving stars of about the same range of initial masses (as for instance ratios of blue to red supergiants, of Wolf-Rayet (WR) to O-type stars). 
Star formation history is not involved either provided we concentrate
on regions of constant star formation ({\it i.e.} constant during at least the last 10-20 millions years). In that case the number ratios of massive stars in two different evolutionary stages is given
by the ratio of the durations of these evolutionary stages\footnote{Let us note that for starburst the situations is different, because the massive star population varies as a function of the age and
depends also on the duration and intensity of the burst of star formation.}. In the following
we briefly recall the main characteristics of the observed populations of Be stars, red and blue supergiants and Wolf-Rayet stars and deduce some consequences for the stellar models.

\subsection{The Be stars}

Be stars are emission line stars. Emission originates in a circumstellar outflowing disk.
How do these disks form~? How long are their lifetimes~? Are they intermittent~? 
Are they Keplerian~? Many of these questions are still subject of lively debate. A point however which seems well accepted is the fact that
the origin of a disk might be connected to the fast
rotation of the star (Pelupessy et al. 2000).
Martayan et al. (2006) showed that the initial velocities
of the Be stars is significantly higher than the initial velocities of the normal B stars,
giving some support to the view that
only stars with a sufficiently high initial velocity can go through a Be star episode.

\begin{figure}
\includegraphics[width=2in,height=2.1in]{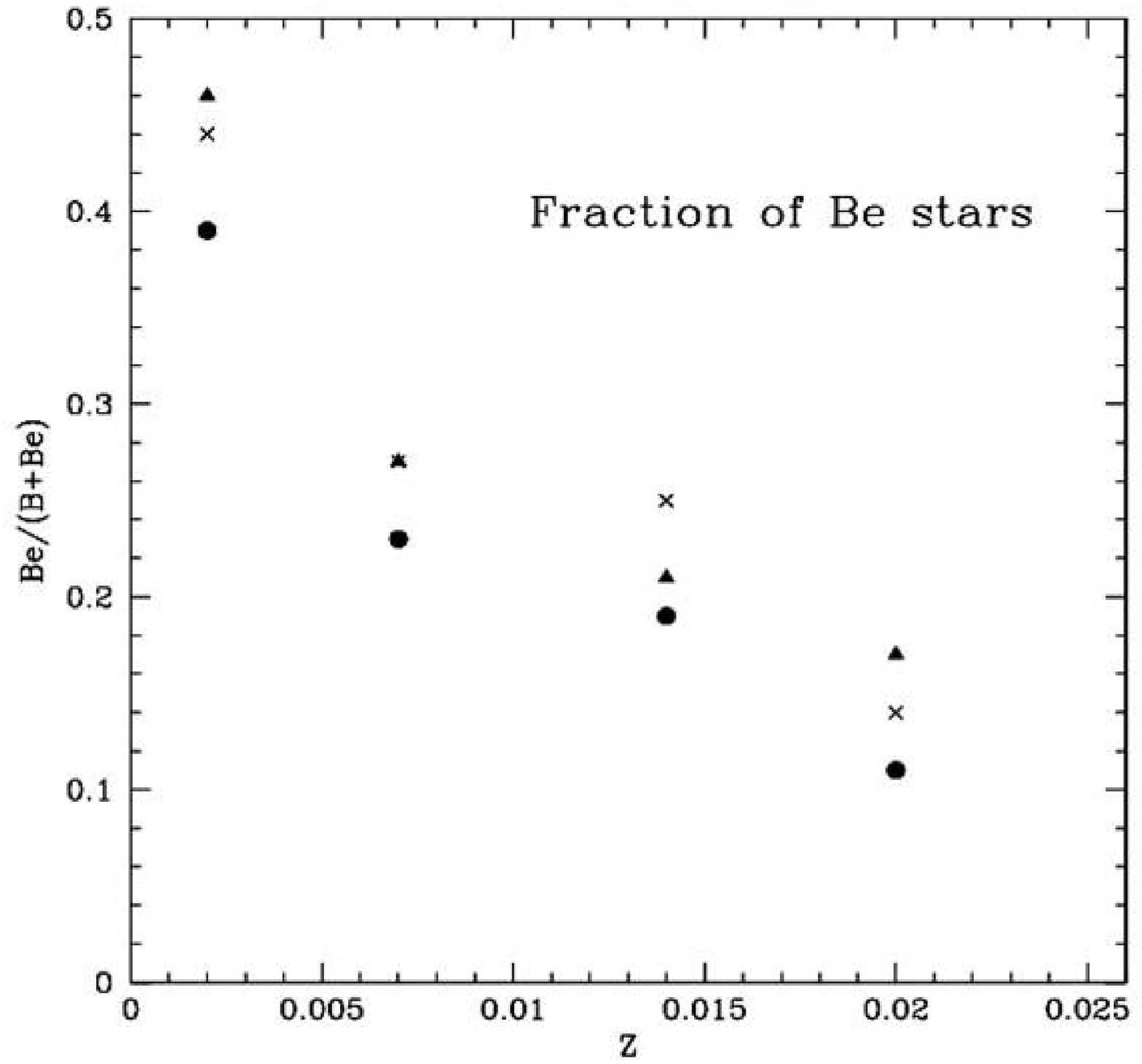}
\hfill
\includegraphics[width=3in,height=2.1in]{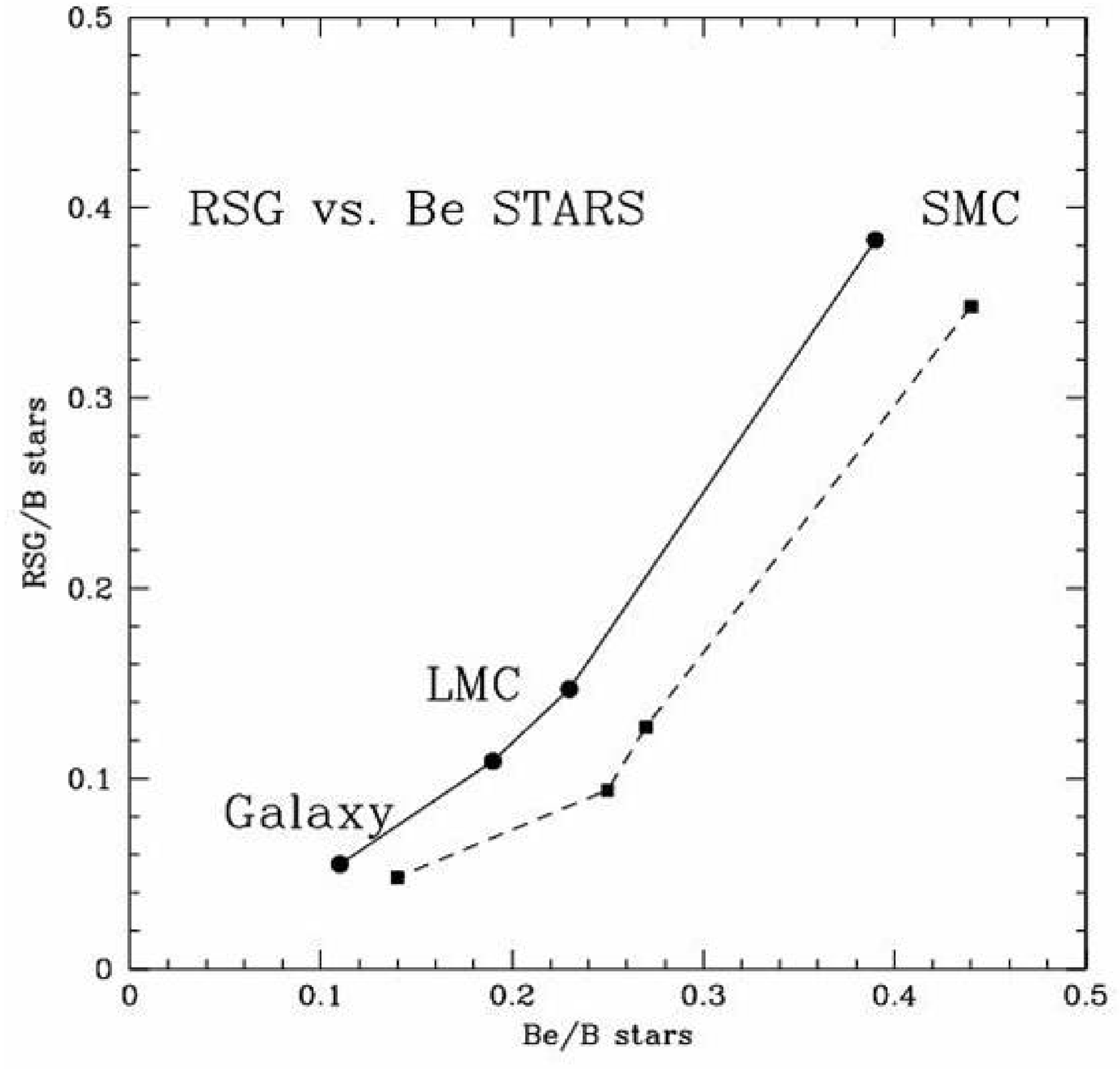}
\caption{{\it Left panel}: Variation as a function of metallicity of the number ratio
of Be stars to B and Be stars (Maeder et al. 1999) 
{\it Right panel}: Correlation between the the number of red supergiants and the number ratio
of Be to B stars.}\label{fig0}
\end{figure}

The population of Be stars varies with the metallicity.
Fig.~\ref{fig0}, {\it left panel}, shows that the number of Be stars with respect to the total number of B stars
(B and Be stars) in cluster with ages between 10 and 25 My (mass at the turn off between
about 9 and 15M$_\odot$) increases with decreasing metallicity (Maeder et al. 1999).
Such a trend has been recently confirmed by Wisniewski \& Bjorkman (2006).
Very interestingly there appears to be a correlation between the frequency of Be stars and that of red supergiants as shown in Fig.~\ref{fig0} {\it right panel}.

These observations indicate that metallicity plays a role in the Be phenomenon and
provides hints on the way surface velocity may evolve differently for stars of different
initial metallicities. The correlation of Be star populations with those of red supergiants
can be seen as an indication that fast rotation not only favors the formation of Be stars
but also that of red supergiants (see below).

\subsection{The supergiants}

\begin{figure}
\includegraphics[width=2.5in,height=2.1in]{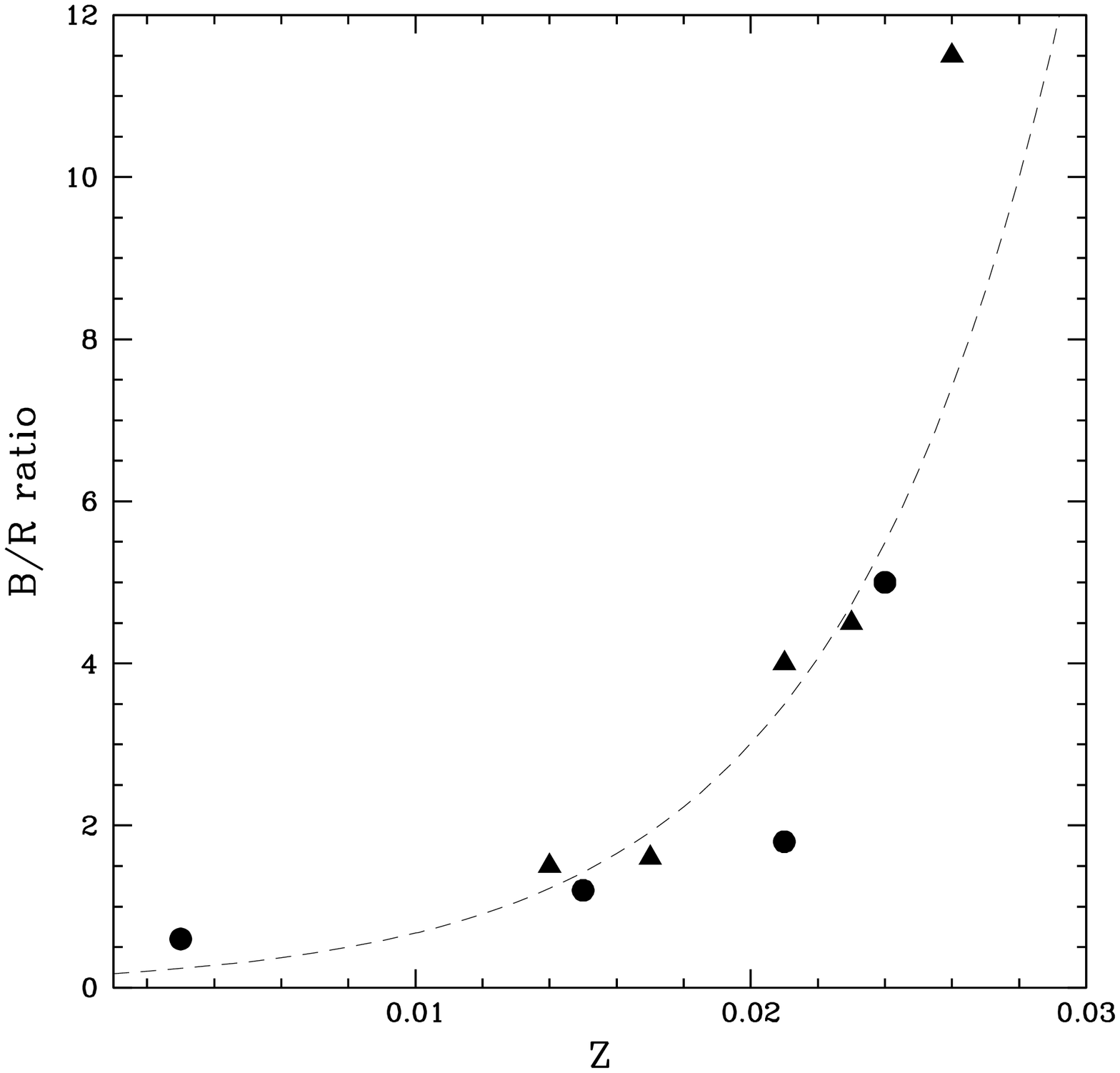}
\hfill
\includegraphics[width=2.5in,height=2.1in]{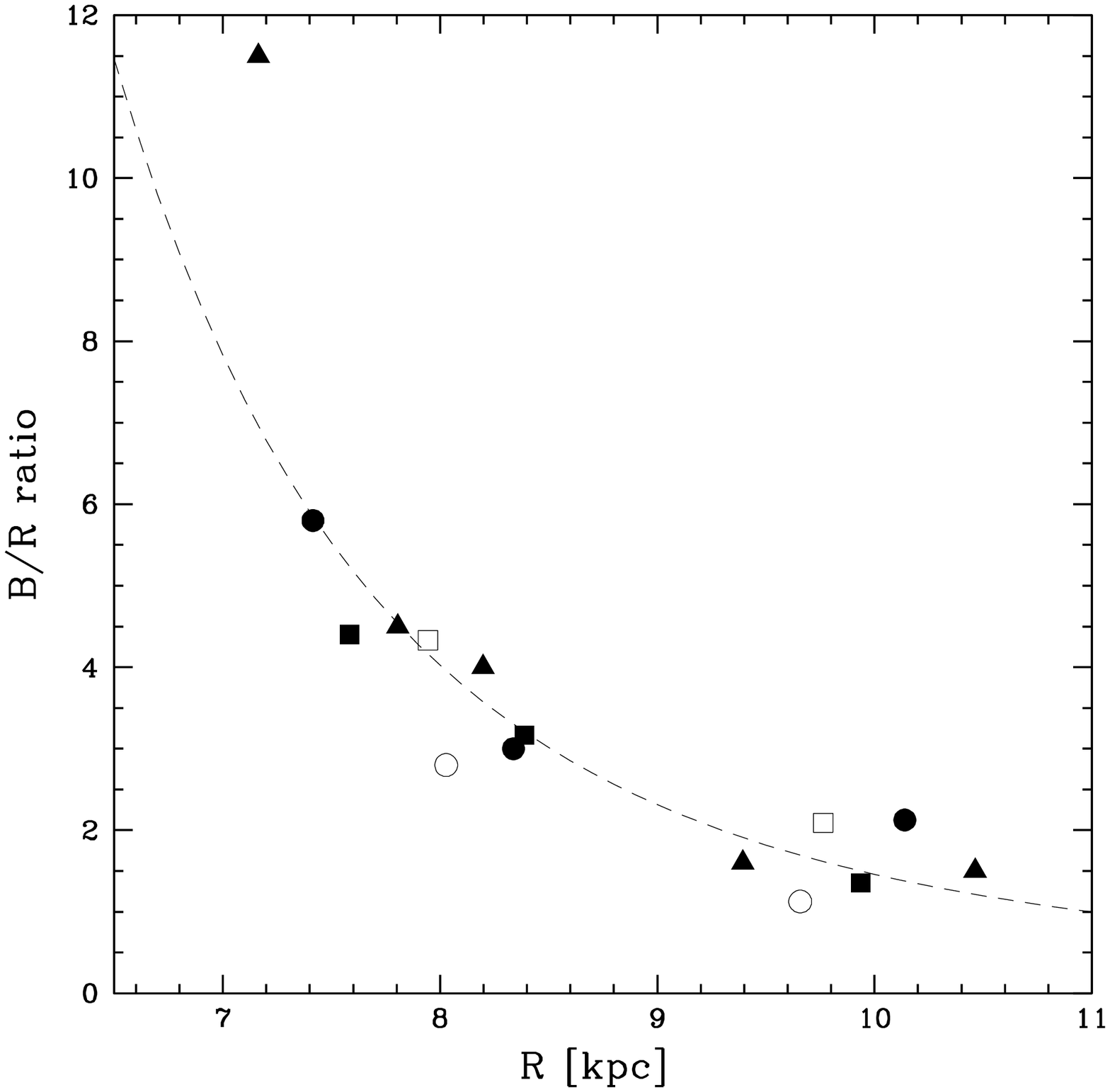}
\caption{{\it Left panel}:  $B/R$ ratio in the Galaxy and the SMC for clusters with $\log age$ between 6.8 and 7.5. The distinction between blue and red
supergiants is based on spectroscopic measurements. The triangles refer to $B$
including O, B and A supergiants.
The dots refer to $B$ including only B supergiants. The dashed curve corresponds to the
fit for $B$ including O, B and A supergiants with $(B/R)_{\odot}=3.0$.
{\it Right panel}: $B/R$ ratio in the Galaxy for different age intervals, with distinction between blue and red
supergiants based on spectroscopic measurements.
$B$ includes O, B and A supergiants. The dashed curve corresponds to the same 
fit as in the left panel. For a detailed description of the figure see Eggenberger et al. (2002)}\label{fig1}
\end{figure}

In Fig.~\ref{fig1}, {\it left panel}, is shown the variation with the metallicity of the number ratio of blue to red supergiants (B/R), and ({\it right panel}) its variation as a function of the galactocentric distance in the Milky Way (Meylan \& Maeder 1982; Eggenberger et al. 2002). Clearly, the B/R ratio increases when the metallicity increases, while standard stellar evolution models ({\it i.e.} models without any extra-mixing processes) predict that this ratio should decrease with increasing metallicity. 
Dohm-Palmer \& Skillman (2002) have
estimated the B/R ratio in the dwarf irregular galaxy Sextans A for various age bins spanning
a range between 20 and 140 My. They find that the observed B/R are lower than those given by standard models by a factor 2, indicating that too few red supergiants are predicted by these models. 

This disagreement implies that no reliable predictions can be made concerning
the nature of the supernova progenitors in different environments, or the populations of supergiants in galaxies. 
The B/R ratio also constitutes an important and sensitive test for stellar evolution models, because it is very sensitive to mass
loss, convection and mixing processes (Langer \& Maeder 1995). 
Thus, the problem of the blue to red supergiant ratio ($B/R$ ratio)
remains one of the most severe problems in stellar evolution.

\subsection{The Wolf-Rayet stars}

Wolf--Rayet stars play a very important role in Astrophysics, as signatures
of star formation in galaxies and starbursts, as injectors of chemical elements and of the 
radioactive isotope $^{26}$Al, as  sources of kinetic energy into the interstellar medium and 
as progenitors of supernovae and, likely, as progenitors of long soft $\gamma$--ray bursts.

Let us recall
some difficulties faced by standard stellar models concerning the WR stars.
  A good agreement between 
the predictions of the stellar models for the WR/O number ratios and the observed 
ones at different metallicities in regions of constant star formation was achieved 
provided the mass loss rates were enhanced by about a factor of two during the MS
and WNL phases (Maeder \& Meynet 1994). This solution, which at that time appeared 
reasonable in view of the uncertainties pertaining the mass loss rates, is no longer
applicable at present. Indeed, the mass loss rates during the WR phase are
reduced by a factor 2 to 3, when account is given to the clumping effects
in the wind (Nugis and Lamers 2000).
Also, the mass loss rates for O--type stars have been substantially revised 
(and in general reduced) by the new results of 
Vink et al. (2001).
In this new context, it is quite clear that with these new mass loss rates
the predicted numbers of WR stars by standard non--rotating models would be much
too low with respect to the observations.

A second difficulty of the standard models with mass loss concerns
the observed number of transition WN/WC stars. These stars show simultaneously some 
nitrogen characteristic of WN stars and some carbon of the further WC stage.
The observed frequency of WN/WC stars among WR stars turns around 4.4 \% (van der Hucht 2001), while
the frequency predicted by the standard models without extra--mixing processes 
are lower by 1--2 orders of magnitude (Maeder \& Meynet 1994).
A third difficulty of the standard models as far as WR stars were concerned was that
there were relatively too many WC stars with respect to WN stars predicted at low metallicity (see the review
by Massey 2003). These difficulties are the signs that some process is missing
in standard models.

\section{Effects of rotation}

Many of the observed characteristics of massive stars require to be explained some extra-mixing mechanism working in their radiative zones. We want to explore here the possibility that this extra-mixing mechanism is triggered by the axial rotation of stars. Rotation appears indeed as the most promising mechanism since massive stars are fast rotators and since many instabilities are triggered by rotation (see Talon this book).




Rotation induces many processes in stellar interior (see the review by Maeder \& Meynet 2000).
In particular, it drives instabilities which transport angular momentum and chemical species.
Assuming that
the star rapidly settles into a state of shellular rotation (constant angular velocity 
at the surface of isobars), the transport equations due to meridional currents and shear instabilities
can be consistently obtained (Zahn 1992). Since the work by J.-P.~Zahn, various improvements have been brought to the
formulas giving the velocity of the meridional currents (Maeder \& Zahn 1998), those of the various diffusive coefficients 
describing the effects of shear turbulence (Maeder 1997; Talon \& Zahn 1997; Maeder 2003; Mathis et al. 2004), as well as the effects of rotation on the mass loss (Owocki et al. 1996; Maeder 1999; Maeder \& Meynet 2000). 

Let us recall a few basic results obtained from rotating stellar models:

1) Angular momentum is mainly transported by the meridional currents. In the outer part
of the radiative envelope these meridional currents transport angular momentum outwards.
During the Main-Sequence phase, the core contracts and the envelope expands. The meridional currents
imposes some coupling between the two, slowing down the core and accelerating the outer layers.
In the outer layers, the velocity of these currents becomes smaller when the density gets higher, {\it i.e.},
for a given initial mass, when the metallicity is lower.

2) The chemical species are mainly transported by shear turbulence (at least in absence of
magnetic fields; when magnetic fields are amplified by differential rotation as in the Tayler-Spruit
dynamo mechanism, Spruit 2002 , the main transport mechanism is meridional circulation, Maeder \& Meynet 2005). 
During the Main-Sequence
this process is responsible for the nitrogen enhancements observed at the surface of OB stars (see e.g. Heap et al. 2006). The shear turbulence is stronger when the gradients of the angular velocity are stronger. Due to point 1 above, the gradients of $\Omega$ are stronger in metal poor stars and thus the mixing of the chemical
elements will be stronger in these stars. This is illustrated on the left panel of Fig.~\ref{rSMC}
(see the tracks for the 9 M$_\odot$ stellar models).
Looking at the 40 M$_\odot$ stellar model in the left part of Fig.~\ref{rSMC}, one sees that the higher metallicity model presents the highest surface enrichments, in striking contrast with the behaviour of
the 9 M$_\odot$ model. This comes from the fact that
the changes occurring at the surface
of the 40 M$_\odot$ are not only due to rotation but also to mass loss which is more efficient at higher $Z$.

Some observations indicate that rotational mixing might be more efficient at lower metallicities (Venn 1999; Venn \& Przybilla 2003). Let us note also that
the efficiency of the mixing will vary from one element to another. If an element is strongly and rapidly built up in the convective core, it will diffuse by rotational mixing more rapidly in the radiative envelope than an element with a smoother gradient between the convective core and the radiative envelope. This explains
why the stellar surface will be more rapidly enriched in nitrogen than in helium. 
Let us add that mixing is more efficient in stars with increasing initial masses and increasing initial velocities.

In addition to these internal transport processes, rotation also modifies the physical properties
of the stellar surface. Indeed the shape of the star is deformed by rotation (a fact which is now put in evidence 
observationally thanks to the interferometry, see Domiciano de Souza et al. 2003). Rotation implies a non-uniform brightness (also now
observed, see e.g. Domiciano de Souza et al. 2005).
The polar regions are brighter than the equatorial ones. This is a consequence of the hydrostatic
and radiative equilibrium (von Zeipel theorem 1924). In addition, as a result of the internal transport processes,
the surface velocity and the surface chemical composition are modified.

\section{Rotation and mass loss}

We can classify the effects of rotation on mass loss in three categories.

\begin{enumerate}
\item The structural effects of rotation.
\item The changes brought by rotation on the radiation driven stellar winds.
\item The mechanical wind induced by rotation at break-up.
\end{enumerate}

Let us now consider in turn these various processes.

\subsection{Structural effects of rotation on mass loss}

\begin{figure}
\includegraphics[width=2.5in,height=2.5in]{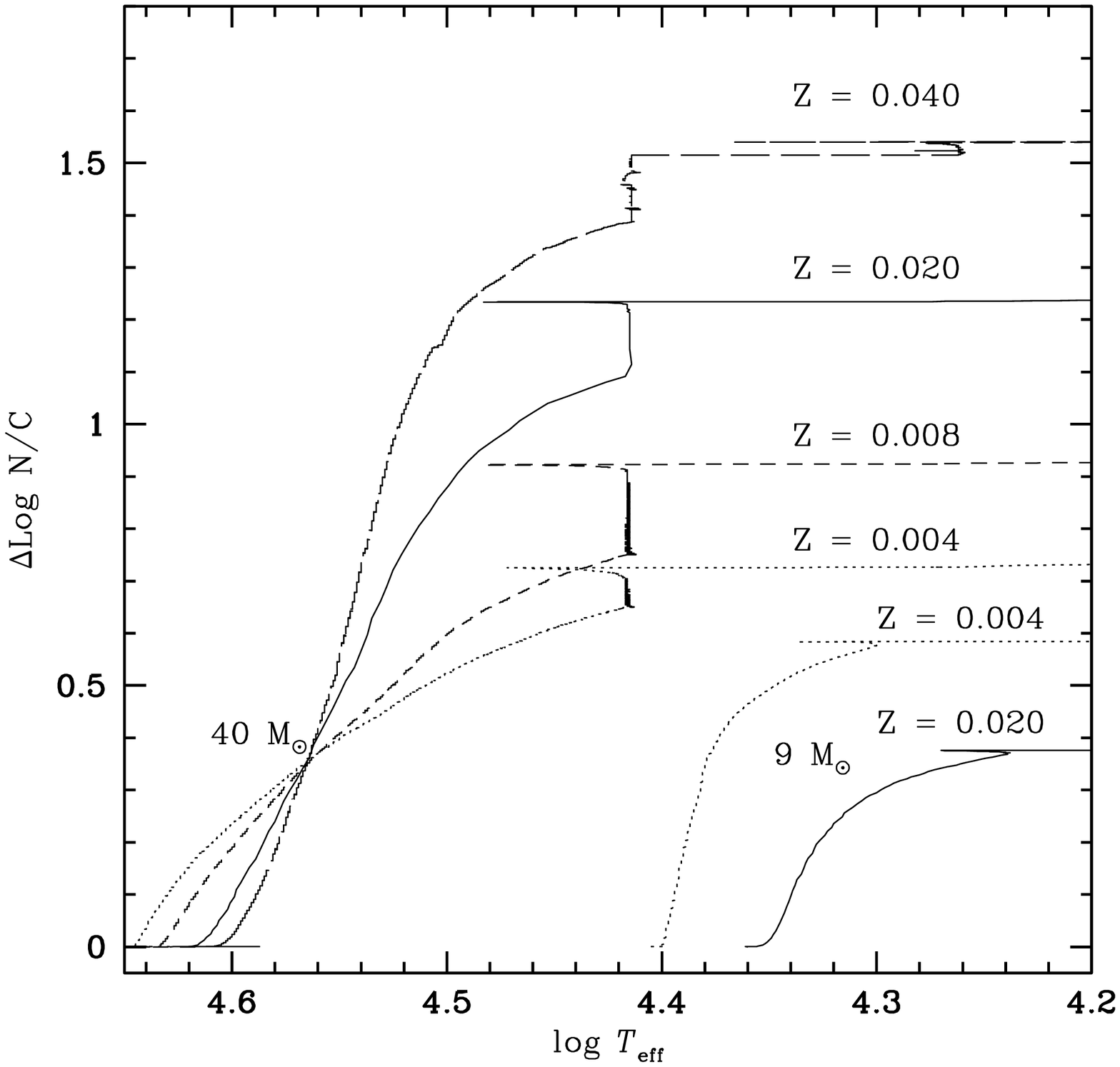}
\hfill
\includegraphics[width=2.5in,height=2.5in]{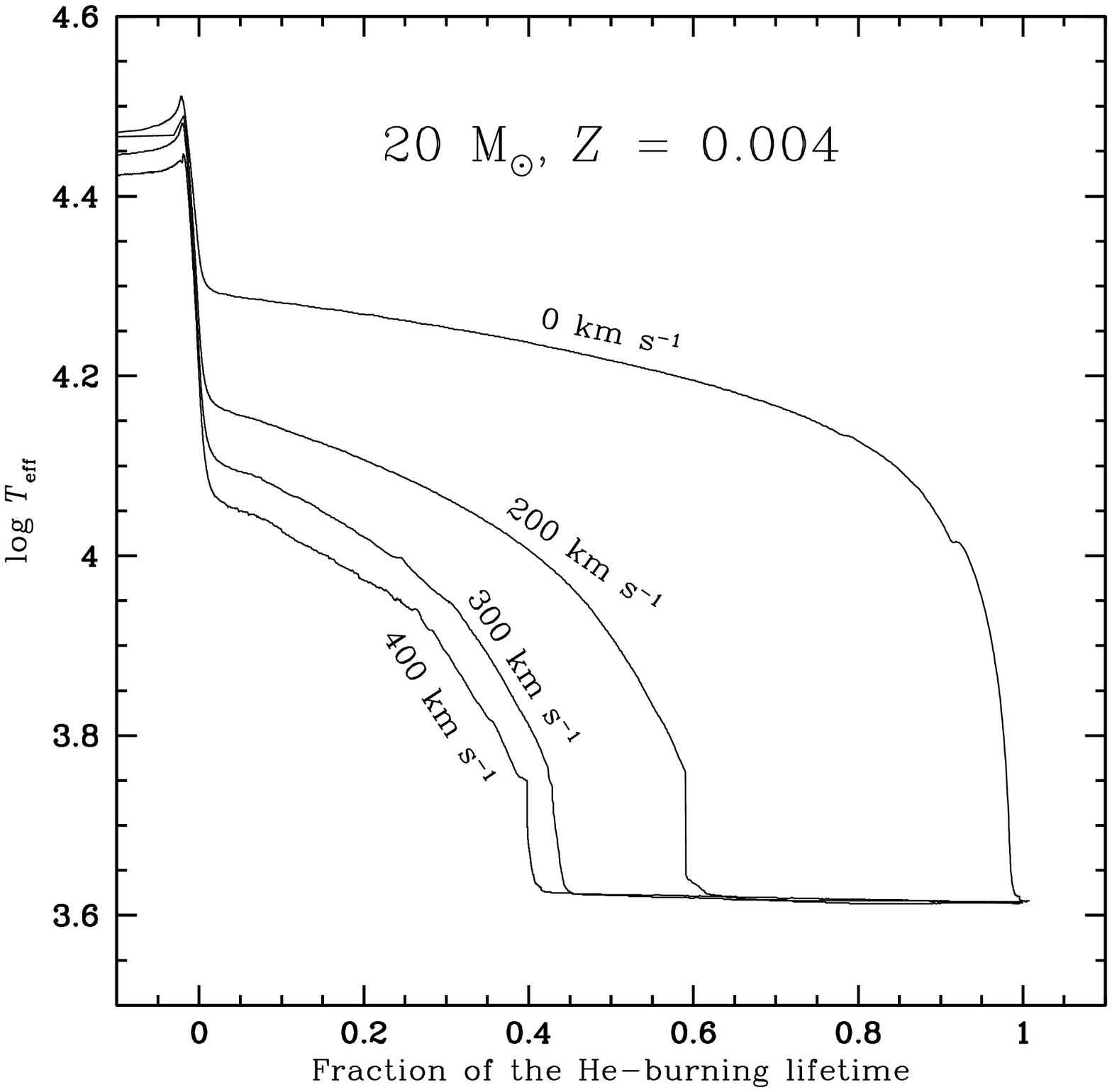}
\caption{{\it Left panel}: Evolution during the MS phase of the N/C ratios (in number) at the surface of rotating stellar models as a function of the effective temperature. 
The differences in N/C ratios are given with respect to the initial values.
{\it Right panel}: Evolution of the $T_{\mathrm{eff}}$
as a function of the fraction of the lifetime spent
in the He--burning phase for 20 M$_\odot$ stars with different
initial velocities.  }
\label{rSMC}
\end{figure}

Rotation, by changing the chemical structure of the star, modifies
its evolution. For instance, moderate rotation at metallicities of the Small Magellanic Cloud (SMC)
favors redward evolution in the Hertzsprung-Russel diagram. This behavior is illustrated in the right panel of Fig.~\ref{rSMC} and can account for the high number of red supergiants observed in the SMC (Maeder \& Meynet 2001),
an observational fact which is not at all reproduced by non-rotating stellar models.

Now it is well known that the mass loss rates are greater
when the star evolves into the red part of the HR diagram, thus in this case, rotation modifies 
the mass loss indirectly, by changing the evolutionary tracks. 
The $\upsilon_{\rm ini}=0$, 200, 300 and 400 km s$^{-1}$ models lose respectively 0.14, 1.40, 1.71 and 1.93 M$_\odot$ during the core He-burning phase (see Table~1 in Maeder \& Meynet 2001). The enhancement of the mass lost reflects the longer lifetimes of the red supergiant phase when velocity increases.
Note that these numbers were obtained assuming that the scaling law between mass loss and metallicity
deduced from stellar wind models for hot stars
applies during the red supergiant phase. If, during this phase, mass loss comes from continuum-opacity driven
wind then the mass-loss rate will not depend on metallicity (see the review by van Loon 2006).
In that case, the redward evolution favored by rotation would have a greater impact on mass loss than
that shown by these computations.

At very high rotation, the star will have a homogeneous evolution and will never become a red supergiant
(Maeder 1987).
In this case, the mass loss will be reduced, although this effect will be somewhat  compensated by two
other processes: first by the fact that the Main-Sequence lifetime will last longer and, second,
by the fact that the star will enter the Wolf-Rayet phase (a phase with high mass loss rates) at an earlier stage of its evolution.

\subsection{Radiation driven stellar winds with rotation}

The effects of rotation on the radiation driven stellar winds 
result from the changes brought by rotation to the stellar surface. They induce changes of the morphologies
of the stellar winds and increase their intensities.

\subsubsection{Stellar wind anisotropies}

\begin{figure}
\resizebox{5cm}{!}{\includegraphics{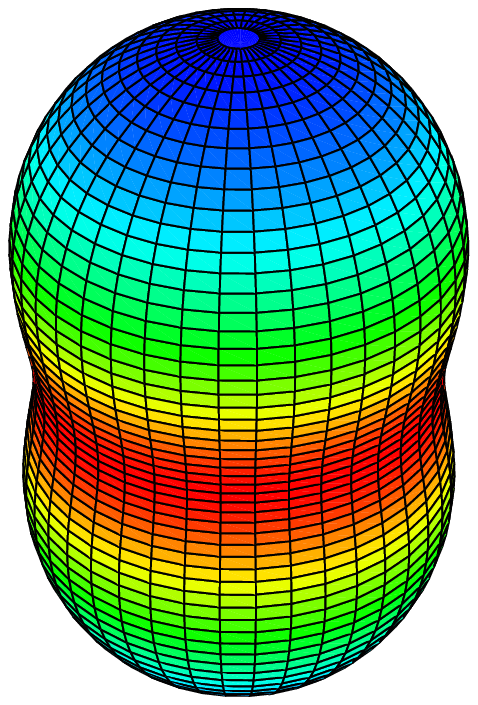}}
\hfill
\includegraphics[width=2.5in,height=2.5in]{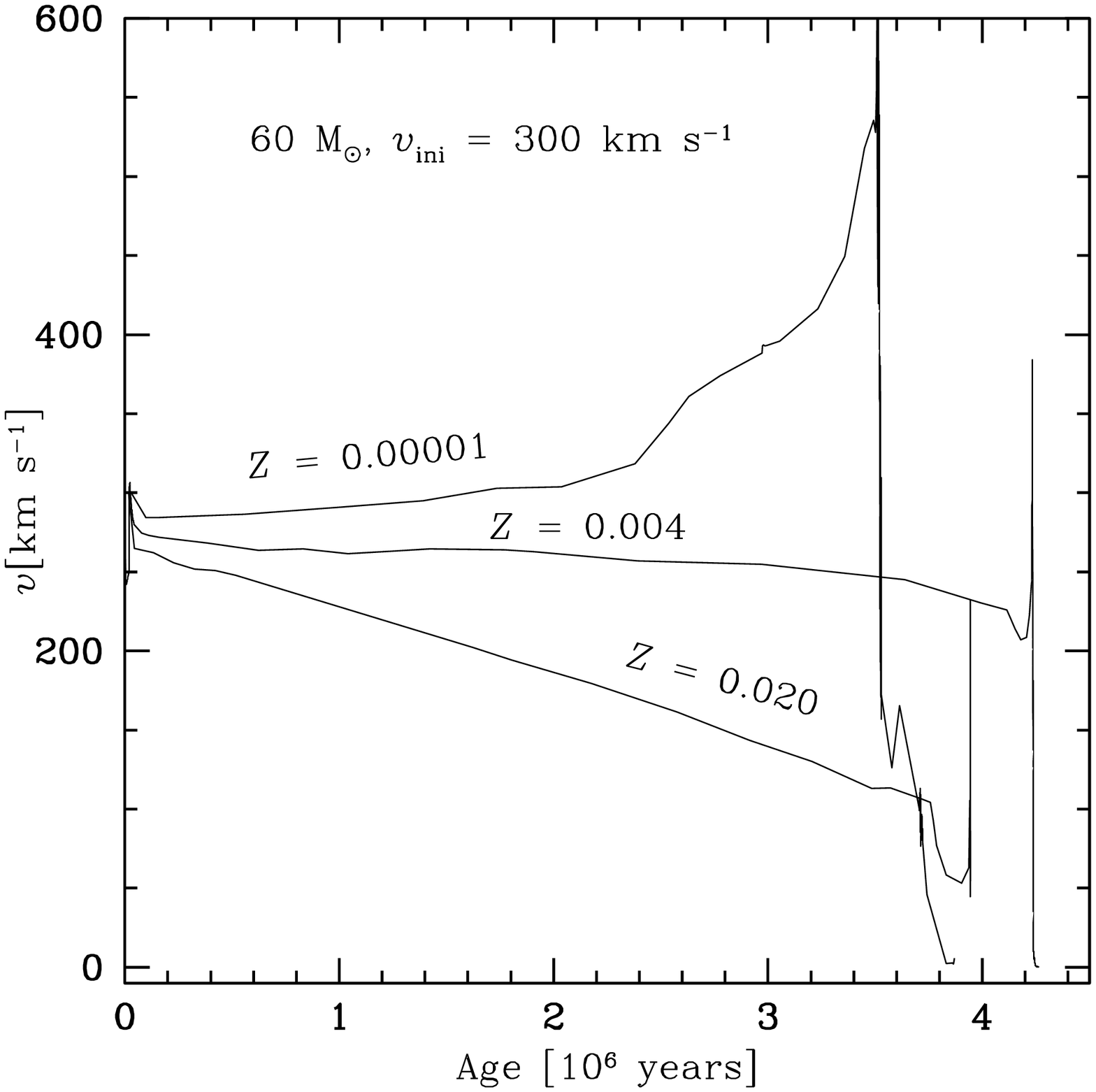}
\caption{{\it Left panel}: Iso-mass loss distribution for a 120 M$_\odot$ star with Log L/L$_\odot$=6.0 and T$_{\rm eff}$ = 30000 K rotating at a fraction 0.8 of critical velocity (figure
from Maeder \& Desjacques 2001).   
{\it Right panel}: Evolution of the surface velocities for a 60 M$_
 {\odot}$ star with 3 different
 initial metallicities. }
\label{ani}
\end{figure}

Naively we would first guess that a rotating
star would lose mass preferentially from the equator, where the effective gravity (gravity decreased
by the effect of the centrifugal force) is lower.
This is probably true when the star reaches the critical limit (i.e. when the equatorial surface
velocity is such that the centrifugal acceleration exactly compensates the gravity), but this is not
correct when the star is not at the critical limit. Indeed as recalled above, a rotating star has a
non uniform surface brightness, and the polar regions are those which have the most powerful radiative 
flux. Thus one expects that the star will lose mass preferentially along the rotational axis. This is
correct for hot stars, for which the dominant source of opacity is electron scattering. In that
case the opacity only depends on the mass fraction of hydrogen and does not depends on other physical quantities such as temperature. Thus rotation induces 
anisotropies of the winds   (Maeder \& Desjacques 2001; Dwarkadas \& Owocki 2002).
This is illustrated in Fig.~\ref{ani}.
Wind anisotropies have consequences for the angular momentum that a star retains in its interior.
Indeed, when mass is lost preferentially along the polar axis, little angular momentum is lost.
This process allows loss of mass without too much loss of angular momentum a process which might
be important in the context of the evolutionary scenarios leading to Gamma Ray Bursts (GRB). Indeed 
in the framework of the collapsar scenario (Woosley 1993), 
one has to accommodate two contradictory requirements: on one side, the progenitor needs to lose mass
in order to have its H and He-rich envelope removed at the time of its explosion, and on the other hand it must have retained sufficient angular momentum in its central region to give birth to a fast
rotating black-hole. Wind anisotropies allow a fast rotating star to lose mass without losing too much angular momentum and can thus be a key factor in the evolution leading to long soft GRB (Meynet
\& Maeder 2007).

\subsubsection{Intensities of the stellar winds}      
      
The quantity of mass lost through radiatively driven stellar winds is enhanced by rotation. This enhancement can occur through two channels: by reducing the effective gravity at the surface of the star, by increasing the opacity of the outer layers through surface metallicity enhancements due to rotational mixing.
      
\begin{itemize}

\item{\it reduction of the effective gravity: } The ratio of the mass loss rate of a star with a surface angular velocity $\Omega$ to that
of a non-rotating star, of the same initial mass, metallicity and lying at the same position in the
HR diagram is given by (Maeder \& Meynet 2000)
      
\begin{equation}
\frac{\dot{M} (\Omega)} {\dot{M} (0)} \simeq
\frac{\left( 1  -\Gamma\right)
^{\frac{1}{\alpha} - 1}}
{\left[ 1 - 
\frac{4}{9} (\frac{v}{v_{\mathrm{crit, 1}}})^2-\Gamma \right]
^{\frac{1}{\alpha} - 1}} \; ,
\end{equation}
\noindent
where $\Gamma$ is the electron scattering opacity for a non--rotating
star with the same mass and luminosity, $\alpha$ is a force multiplier (Lamers et al. 1995). 
The enhancement factor remains modest for stars with luminosity sufficiently far away from the
Eddington limit (Maeder \& Meynet 2000). Typically, $\frac{\dot{M} (\Omega)} {\dot{M} (0)} \simeq 1.5$ for main-sequence B--stars.
In that case, when the surface velocity approaches the critical limit, the effective
gravity decreases and the radiative flux also decreases. Thus the matter becomes less bound
when, at the same time, the radiative forces become also weaker. 
When the stellar luminosity approaches the Eddington limit, the mass loss increases can be much greater, reaching orders of magnitude.
This comes from the fact that rotation lowers the maximum luminosity or the Eddington luminosity of a star.  Thus it may happen that for a velocity still 
far from the classical critical limit, the 
rotationally decreased maximum luminosity becomes equal to the actual luminosity of the star. 
In that case, strong mass loss ensues and the star is said to have reached
the $\Omega\Gamma$ limit (Maeder \& Meynet 2000).

\item {\it Effects due to rotational mixing: }
During the core helium burning phase, at low metallicity,
the surface may be strongly enriched in both H-burning and He-burning products, {\it i.e.} mainly in nitrogen, carbon and oxygen. Nitrogen is produced by transformation of the carbon and oxygen produced in the He-burning core and which have diffused by rotational mixing in the H-burning shell (Meynet \& Maeder 2002). Part of the carbon and oxygen produced in the He-core also diffuses up to the surface. Thus at the surface, one obtains very high value of the CNO elements. For instance a 60 M$_\odot$ with Z=$10^{-8}$ and $\upsilon_{\rm ini}=800$ km s$^{-1}$ has, at the end of its evolution, a CNO content at the surface equivalent to 1 million times its initial metallicity! In the present models, we have applied the usual scaling laws linking the surface metallicity
to the mass loss rates (see Eq.~1). In that case, one obtains that the star loses due to this process
more than half of its initial mass (see Table 3).

\end{itemize}
      
\subsection{Mechanical winds induced by rotation}      
      
As recalled above, during the Main-Sequence phase the core contracts
and the envelope expands. In case of local conservation of the angular momentum, the core would thus
spin faster and faster while the envelope would slow down. In that case, it can be easily shown that the surface velocity would evolve away from the critical velocity (see e.g. Meynet \& Maeder 2006). 
In models with shellular rotation however
an important coupling between the core and the envelope is established through the action of the
meridional currents. As a net result, angular momentum is brought from the inner regions to the outer ones. Thus, would the star lose no mass by radiation driven stellar winds (as is the case at low Z), one expects that the surface velocity
would increase with time and would approach the critical limit. In contrast, 
when radiation driven stellar winds are important, the timescale for removing mass 
and angular momentum at the surface
is shorter than the timescale for accelerating the outer layers by the above process and the surface velocity decreases as a function of time. It evolves away from the critical limit. 
Thus, an interesting situation occurs: when the star loses
little mass by radiation driven stellar winds, it has more chance to lose mass by a mechanical wind. On the other hand, when the star loses mass at a high rate by
radiation driven mass loss then it has no chance to reach the critical limit and thus to undergo a 
mechanical wind. We discuss further below the possible importance of this mechanical wind.

\subsection{Discussion}  

At this point it is interesting to discuss three aspects of the various effects described above. First what
are the main uncertainties affecting them?  Second, what are their relative importance? 

\subsubsection{Uncertainties}

In addition to the usual uncertainties affecting the radiation driven mass loss rates, the above processes 
poses three additional problems:
\begin{enumerate}

\item {\it What does happen when the CNO content of the surface increases by six orders of magnitude as was obtained
in the 60 M$_\odot$ model described above?} Can we apply the usual scaling law between Z and the mass losses?
This is what we have done in our models (using $\alpha=0.5$), but of course this should be studied in more details by stellar winds models. For instance, for WR stars, Vink \& de Koter (2005) have shown that at $Z=Z_\odot/30$, 60\% of the driving is due to CNO elements and only 10\% to Fe.
Here the high CNO surface enhancements result
from rotational mixing which enrich the radiative outer region of the star in these elements, but also from the fact that the star evolves to the red part of the HR diagram, making an outer convective zone to appear. This
convective zone plays an essential role in dredging up the CNO elements at the surface. Thus what is needed
here is the effects on the stellar winds of CNO enhancements in a somewhat red part of the HR diagram
(typical effective temperatures of the order of Log T$_{\rm eff}\sim$3.8).

\item {\it Can stars reach the critical limit?} For instance, Bodenheimer \& Ostriker (1970)
obtain that during pre-main sequence evolution of rapidly rotating massive stars, ``equatorial mass loss'' or ``rotational mass ejection'' never occur (see also Bodenheimer \& Ostriker 1973). In these models the condition of zero effective gravity is never reached.  However, these authors studied pre-main sequence evolution (while here MS evolution is considered) and made different hypotheses on the
transport mechanisms than in the present work. Since they were interested in the radiative contraction
phase, they correctly supposed that ``the various instabilities and currents which transport angular momentum
have characteristic times much longer than the radiative-contraction time''. This is no longer the case 
for the Main-Sequence phase. In our models, we consistently accounted for the transport of the angular
momentum by the meridional currents and the shear instabilities.
A detailed account of the transport mechanisms shows that they are never able to prevent the star from reaching
the critical velocity.
Another difference between the approach in the work of Bodenheimer \& Ostriker (1970) and ours is that these authors consider another distribution of the angular velocity than in our models. They supposed constant $\Omega$
on cylindrical surface, while here we adopted, as imposed by the theory of Zahn (1992), a ``shellular rotation law''. They resolved the Poisson equation for the gravitational potential, while here we adopted the Roche model. Let us note that the Roche approximation appears justified in the present case, since only the outer layers, containing little mass, are approaching the critical limit. The majority of the stellar mass has a rotation rate much below the critical limit and is thus not strongly deformed by rotation. Thus these differences probably explain why in our models we reach situations where the effective gravity becomes zero.

\item {\it What does happen when the surface velocity reaches the critical limit?} 
Let us first note that when the surface reaches the critical velocity, the energy which is still needed to
make equatorial matter to escape from the potential well of the star is still important. This is because the gravity of the system continues of course to be effective all along the path from the surface to the infinity
and needs to be overcome.
The escape velocity of a piece of material of mass $m$ at the equator
of a body of mass $M$, radius $R$, rotating at the critical velocity, is
\begin{equation}
{1 \over 2}m \upsilon_{\rm crit}^2+{1 \over 2}m \upsilon_{\rm esc}^2-{GMm \over R}=0,
\end{equation}
one obtains, using $\upsilon_{\rm crit}^2={GM/R}$  that the escape velocity is 
simply reduced by a factor $1/\sqrt{2}=0.71$ with respect to the escape velocity from a non-rotating body
\footnote{We suppose here that
the vector $\upsilon_{\rm esc}$ is normal to the direction of the vector $\upsilon_{\rm crit}$.}.
Thus the reduction is rather limited and one can wonder if matter will be really lost.
A way to overcome this difficulty is to consider the fact that, at the critical limit, the matter will
be launched into a keplerian orbit around the star.
Thus, probably, when the star reaches
the critical limit an equatorial disk is formed like for instance around Be stars. 
This disk will probably dissipate by radiative effects and thus the material will be lost by the star.

Practically, in the present models, we remove the supercritical layers. This removal of material allows the outer layers to become again subcritical until secular evolution brings again the surface near the critical limit (see Meynet et al. 2006 for more details
on this process). Secular evolution during the Main-Sequence phase triggers two counteracting effects: on one side, the stellar surface expands. Local conservation of the angular momentum makes the surface to slow down and the surface velocity to evolve away from the critical limit. On the other
hand, meridional circulation continuously brings angular momentum to the surface and accelerates the outer layers. This last effect in general overcomes the first one and the star rapidly reach again the critical limit. How much mass is lost by this process?
As seen above, the two above processes will maintain the star near the critical limit for most of the time.
In the models, we adopt the mass loss rate required 
to maintain the star at about 95-98\% of the critical limit. Such a mass loss rate is imposed  as long as the secular evolution brings back the star near the critical limit. In general, during the
Main-Sequence phase, once the critical limit is reached, the star remains near this limit for the rest
of the Main-Sequence phase. At the end of the Main-Sequence phase, evolution speeds up and
the local conservation of the angular momentum overcomes the effects due to meridional currents, the star
evolves away from the critical limit and the imposed ``critical'' mass loss is turned off.

\end{enumerate}

\subsubsection{Importance of the various effects on mass loss induced by rotation}

The processes which are the most important for metal poor stars are the reaching of the critical limit
(both the classical limit and the $\Omega\Gamma$-limit) and the increase of the surface metallicity by
the concomitant effect of rotational diffusion and dredging-up by an outer convective zone.

In order to quantify the importance of the various effects discussed above, we compare in Table~\ref{Mdot} four
60 M$_\odot$  stellar models with an initial velocity of 800 km s$^{-1}$ at four different metallicities, $Z=0$
(Ekstroem et al. 2005), 10$^{-8}$,
10$^{-5}$ (Meynet et al 2006) and $10^{-3}$ (Decressin et al. 2007) and we give the mass lost during the MS and the post MS (PMS) phases. The mass lost by non-rotating models is also given.

\begin{table}
\begin{center}
\begin{tabular}{|l|l|l|l|}
\hline
           &                            &                       &                         \\
Z          & $\Omega/\Omega_{\rm crit}$ & $\Delta$ M$_{\rm MS}$ & $\Delta$ M$_{\rm PMS}$  \\
           &                            &                       &                         \\
\hline 
0          & 0                          & 0                     & 0.0013 (0)              \\
0          & 0.71                       & 2.42                  & 0.27   (0)              \\
           &                            &                       &                         \\
10$^{-8}$  & 0                          & 0.18                  & 0.09   (0)              \\
10$^{-8}$  & 0.77                       & 2.38                  & 33.80  (0.85)           \\
           &                            &                       &                         \\
10$^{-5}$  & 0                          & 0.21                  & 0.22   (0)              \\
10$^{-5}$  & 0.90                       & 6.15                  & 16.94  (0)              \\
           &                            &                       &                         \\                   0.0005     & 0                          & 0.78                  & 13.29  (0)              \\
0.0005     & 0.94                       & 20.96                 & 21.79  (17.15)          \\
           &                            &                       &                         \\            
\hline
\end{tabular}
\caption{Mass lost in solar masses by 60 M$_\odot$ non-rotating and rotating models at different
metallicities during the MS and the post MS phases. The number in parenthesis in the last column
indicates the mass lost during the WR phase. See text for the references of the stellar models.}
\label{Mdot} 
\end{center}
\end{table}

From Table~\ref{Mdot}, we first note that a given value of the initial velocity (here 800 km s$^{-1}$)
corresponds to lower value of $\Omega/\Omega_{\rm crit}$ at lower metallicity. This is a consequence
of the fact that stars are more compact at low $Z$. Would we have kept $\Omega/\Omega_{\rm crit}$
constant one would have higher values of $\upsilon_{\rm ini}$ at low Z. 

During the MS phase, we see that the non-rotating models lose nearly no mass. 
The rotating models, on the other hands, lose some mass when reaching the critical limit.
For the Pop III star the critical limit
is reached when the mass fraction of hydrogen at the center, $X_c$, is 0.35. For the models at Z= $10^{-8}$,
$10^{-5}$ and 0.0005, the critical limit is reached respectively when $X_c$ is equal to 0.40, 0.56 and
0.65. Thus at higher metallicity, the critical limit is reached earlier.
This behavior comes from two facts: first keeping $\upsilon_{\rm ini}$ constant implies higher
$\Omega/\Omega_{\rm crit}$ at higher $Z$, then, meridional currents, which accelerate the
outer layers are more rapid at higher metallicities.

The mass lost after the Main-Sequence phase remains very modest for non-rotating stars, except for the model
at $Z=0.0005$. For the rotating models, except in the case of the Pop III models, all models lose great
amounts of material. In the case of the models with $Z=10^{-8}$ and $10^{-5}$, the main effect responsible for
the huge mass loss is the surface enrichments in CNO elements. In the case of the $Z=0.0005$, no such effect is observed, however the star, as a result of the high mass loss during the MS phase and also
due to rotational mixing, has a long WR phase, during which most of the mass is lost. The Pop III model
on the other hand loses little amount of mass during the post-MS phase. This comes from the fact that
the star evolves only at the very end of its evolution in the red part of the HR diagram, preventing thus an efficient dredging up of the CNO elements at the surface. Thus the surface enhancements remain modest and occur during a too short phase for having an important impact on mass loss. On the other hand, it would be interesting to compute models with higher initial values of  $\Omega/\Omega_{\rm crit}$.

As a general conclusion, we see that the quantity of mass lost very much depends on rotation in metal poor regions.
Moreover, the lost material is enriched in new synthesized elements like helium, carbon, nitrogen and oxygen and thus will participate to the chemical evolution of the interstellar medium. 
Short comments on this point are made below.

\section{Interesting consequences of rotating models}

In this section we briefly resume the main results deduced from massive star rotating models concerning the evolutionary tracks, lifetimes, the surface abundances and the massive star populations. In the next paragraph we focus on nucleosynthetic aspects. In table~\ref{t1}, recent
grids of rotating massive star models (without magnetic fields) are indicated. Only the main velocity range is indicated in each case. For some metallicities, more than one grid are available.
Note that the grids differ by the mass loss, the overshooting parameter, the expression
of the shear diffusion coefficients. We shall not discuss here in details these models but underline
some of their important characteristics.

\begin{table}
\caption{Grids of rotating massive star models.}
\begin{center}
\begin{tabular}{lcccr}
 & & & &\\
Reference & Initial Masses &  $Z$  & Initial & Models \\ 
          &    M$_\odot$   &       & rotation& computed \\
          &                &       & km/s    & until end of            \\
 & & & &\\
Hirschi 2007           & 9 - 85   &  10$^{-8}$ & 800       & Si-b.       \\
Meynet \& Maeder 2002  & 2 - 60   &  0.00001   & 0 \& 300  & C-b.E-AGB \\
Decressin et al. 2007  & 20 - 120 & 0.0005     & 800       & He-b.        \\
Meynet \& Maeder 2005  & 30 - 120 & 0.004      & 300       & He-b.        \\
Maeder \& Meynet 2001  & 9 - 60   & 0.004      & 0 \& 300  & He-b.        \\
Meynet \& Maeder 2005  & 30 - 120 & 0.008      & 300       & He-b.        \\
Heger et al. 2000      & 8 - 25   & 0.020      & 0 \& 210  & core col.        \\
Meynet \& Maeder 2000  & 9 - 120  & 0.020      & 0 \& 300  & He-b.        \\
Meynet \& Maeder 2003  & 9 - 120  & 0.020      & 0 \& 300  & He-b.        \\
Hirschi et al. 2004    & 12 - 60  & 0.020      & 0 \& 300  & Si-b.        \\ 
Meynet \& Maeder 2005  & 20 - 120 & 0.040      & 0 \& 300  & He-b.       \\
 & & & & \\
\end{tabular}
\end{center}
\label{t1}
\end{table}

During the core H-burning phase phase, the main effects of rotation are the following:
\begin{itemize}
\item Rotating tracks are over luminous and in general redder than non-rotating ones at the end of the MS phase (see e.g. discussion in Meynet \& Maeder 2000). Thus the estimate of the initial mass of a star obtained by looking for the evolutionary track going through its observed position in the HR diagram would overestimate the mass of the star if rotation is not accounted for. 
\item Rotation increases  the MS lifetime with respect to non--rotating models (up to
 about 40 \% for the most massive stars and for very fast rotation). The values assigned from isochrones with an average rotation velocity  typically lead to ages 25\% larger than without rotation.
\item The rotating models account for the observed changes of the surface abundances for OB main sequence  stars (for comparison with surface boron abundance discussed in Sect.~3.1 see Mendel et al. 2006).  
\item Steeper gradients of internal rotation $\Omega$ are built at lower $Z$.
The steeper $\Omega$--gradient at lower $Z$  favors mixing. This effect might be responsible for the higher
surface enrichments observed at the surface of SMC A-type supergiants compared to galactic ones
(Venn 1999; Venn \& Przybilla 2003).
There are 2 reasons for the steeper $\Omega$--gradients. 
One is the higher compactness of the 
star at lower $Z$. The second one is that at lower $Z$, the density of the outer layers
is higher, thus the meridional currents are slowlier. This produces less outward transport of angular momentum and 
contributes to steepen 
the $\Omega$--gradient.
\item  At lower $Z$, rotating stars more easily reach break--up velocities and may
stay at break-up for a substantial fraction of the MS phase. This has interesting consequences
for explaining the Be phenomenon (see Meynet et al. 2007).
\end{itemize}

Rotation has also an impact on the more advanced stellar evolution phases. It favors the redward
evolution in the HR diagram. This particularly striking at low metallicity where models with rotation
predict a high number of red supergiants as required by the observation (Maeder \& Meynet 2001).
Models with rotation can also account for the observed variation of the number of WR to O-type stars
with the metallicity, while non-rotating models predict much too low values for these ratios
when mass loss rates accounting for clumping effect are used (Meynet \& Maeder 2005). Also the rotational models well
reproduce the WN/WC ratio at low metallicity, the observed fraction of WR stars in the transition stage WN/WC and the variation with the metallicity of the number ratios of type Ibc to type II supernovae (see more detailed discussion in Meynet \& Maeder 2003, 2005). Rotating models however face some difficulties in reproducing the observed ratio of WN to WC stars at high metallicity, there is however some hope to resolve this question by adopting the more
recent metallicity dependent mass loss rates for the WR stars (Eldridge \& Vink 2006). Rotating
models would also predict too rapidly rotating young pulsars (Woosley \& Heger 2006b). This may
indicate that the angular momentum is evacuated more efficiently from the central regions than
in the above rotating models, either during the hydrostatic phases of the evolution, at the time
of the supernova explosion or during the very early phases of the pulsar's evolution.
Except for these two difficulties (WN/WC ratio at high metallicity and pulsars rotation rate),
present models with rotation greatly improves on the whole  the correspondence between theory and
observation.
  
\section{Consequences for nucleosynthesis}

Let us begin this section by recalling a few orders of magnitudes. When single stars form with a Salpeter's IMF,
about 14\% of the mass locked into stars consists of massive stars, {\it i.e.} with masses greater than 8 M$_\odot$, 25\%  are locked into stars of masses between 1 and 8 M$_\odot$ and 61\% in stars
with masses between 0.1 and 1 M$_\odot$. When all the stars with masses above 8 M$_\odot$ have died,
about 13\% of the mass initially locked into stars is ejected by massive stars (1\% remains locked into black holes or neutron stars). The intermediate mass stars (from 1 to 8 M$_\odot$) eject about 18.5\% of the mass initially locked into stars (6.5\% remains locked in white dwarfs).

Figure 6 shows for various models at different metallicities the fraction of the mass in stars which is
eventually ejected under the form of new elements. The part ejected by stellar winds is distinguished from that
ejected at the time of the supernova explosion. One sees that on the whole, the fraction of the mass initially locked into stars and transformed into new elements by the massive stars does not depend too much on the model or on the metallicity. All the results are comprised between 3.5 and 4.5\%. 

\begin{figure}
\includegraphics[width=3.4in,height=2.0in]{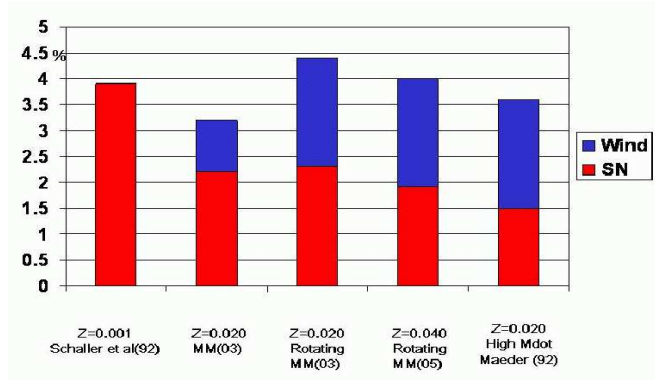}
\caption{Mass of new elements ejected by stars more massive than 8 M$_\odot$
per mass in stars given by different stellar models. A Salpeter IMF has been used. The labels MM03 and MM05
are for respectively Meynet \& Maeder 2003 and Meynet \& Maeder 2005.}\label{fig6}
\end{figure}

One notes however that more variations appear when one looks at the proportions of these new elements ejected
by the winds and the supernova explosion. Indeed at higher metallicity a greater part of the new elements synthesized by stars are ejected by stellar winds
as can be see in Fig~\ref{fig6}. Typically the non-rotating stellar models of Schaller et al. (1992)
predict that a stellar population at $Z=0.001$ ejects in the form of new elements
during the supernova explosion a little less than 4\% of the total mass used to form the stars. Models at solar metallicity and higher eject
about half of their new elements at the time of the supernova explosion and the rest through stellar winds.
This may have a great influence on the final yields as has been shown by Maeder (1992). Indeed, mass loss removes matter at earlier evolutionary stages. The ejected matter has therefore been partially processed by nuclear burning and has a chemical composition different from the one it would have if the matter would have remained locked in the star.

This effect is responsible for many specific enrichments at high metallicity: for instance, because
of this effect, massive star
models are expected to be stronger sources of $^{4}$He, $^{12}$C, $^{22}$Ne, $^{26}$Al and to be less 
important sources of $^{16}$O at high metallicity than at low metallicity.

Since, as seen above, the contribution of stars may vary a lot for some elements depending on the initial metallicity, we divide here the discussion into three subsections, the cases of solar, higher
and lower than solar metallicity. As we shall see, very roughly rotation appears as a key factor at low metallicity, while mass loss by stellar winds is clearly the dominant factor at high metallicity.

\subsection{At solar metallicity}

\begin{figure}
\includegraphics[width=2.5in,height=2.5in]{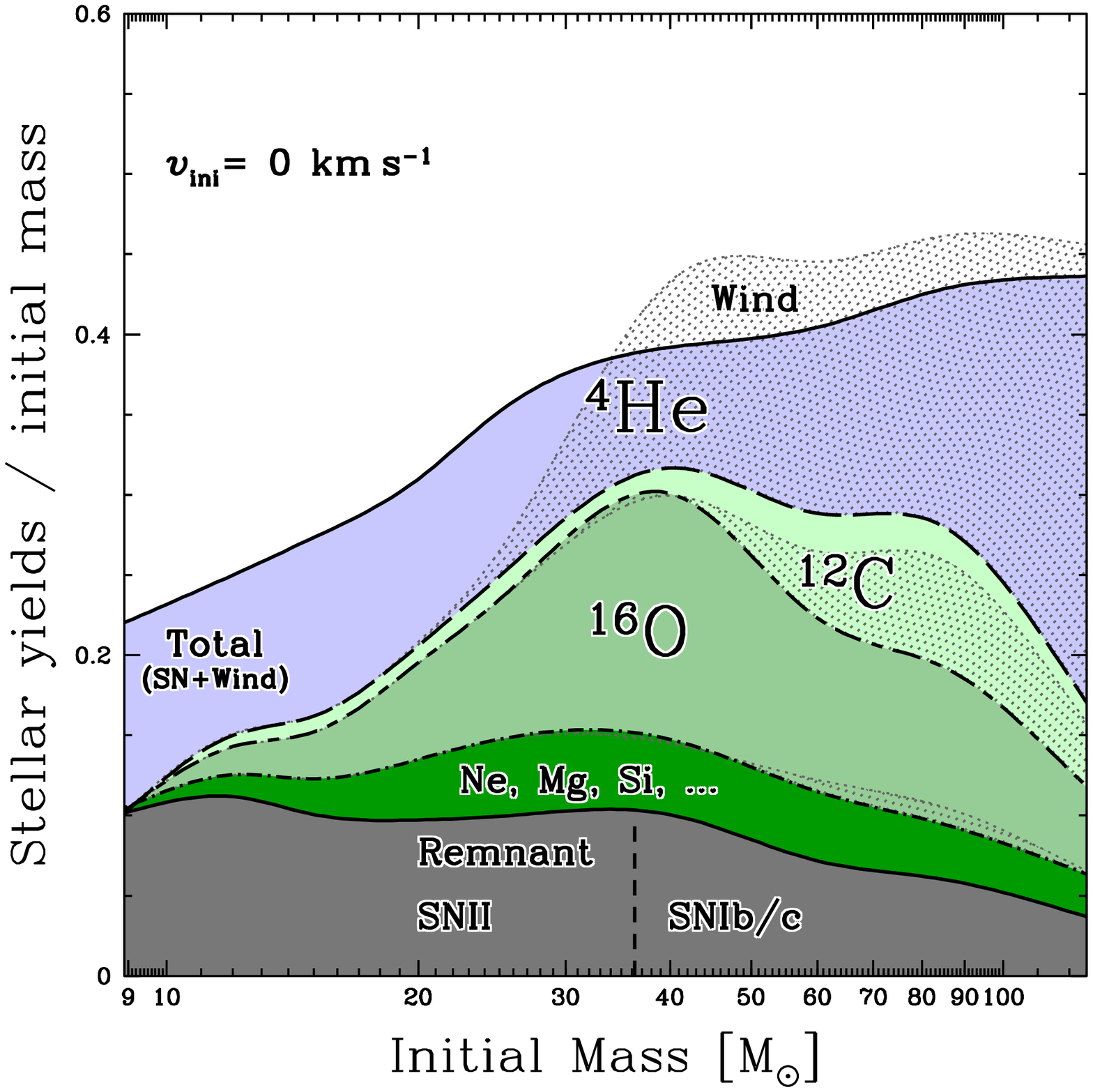}
\hfill
\includegraphics[width=2.5in,height=2.5in]{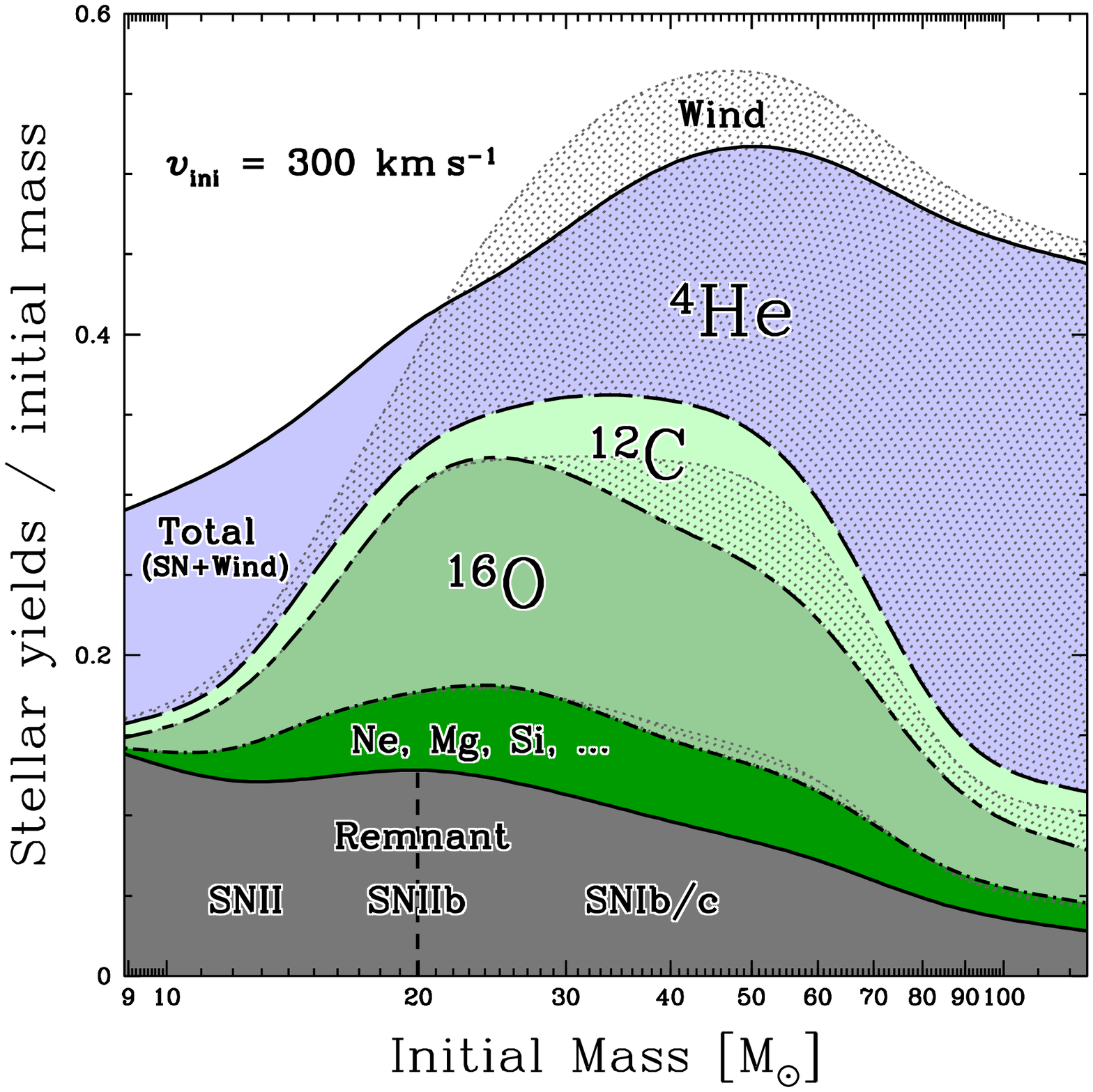}
\caption{{\it Left panel}:  The chemical yields  for models without rotation.
{\it Right panel}: The chemical yields
for models with $\upsilon_{\mathrm{ini}}= 300$ km/s (Hirschi et al. 2005).}
\label{yh}
\end{figure}

\begin{figure}
\includegraphics[width=2.5in,height=2.5in]{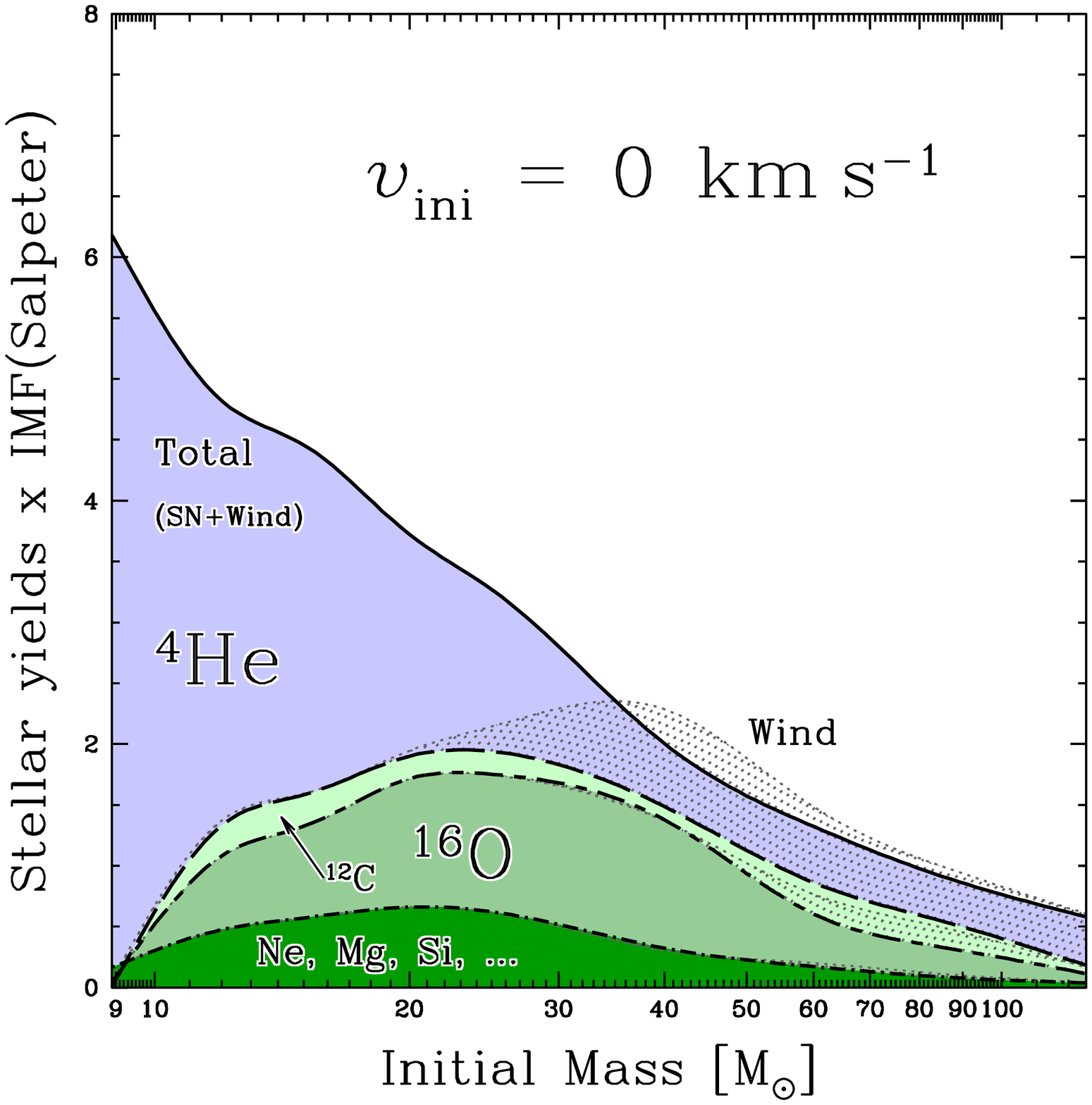}
\hfill
\includegraphics[width=2.5in,height=2.5in]{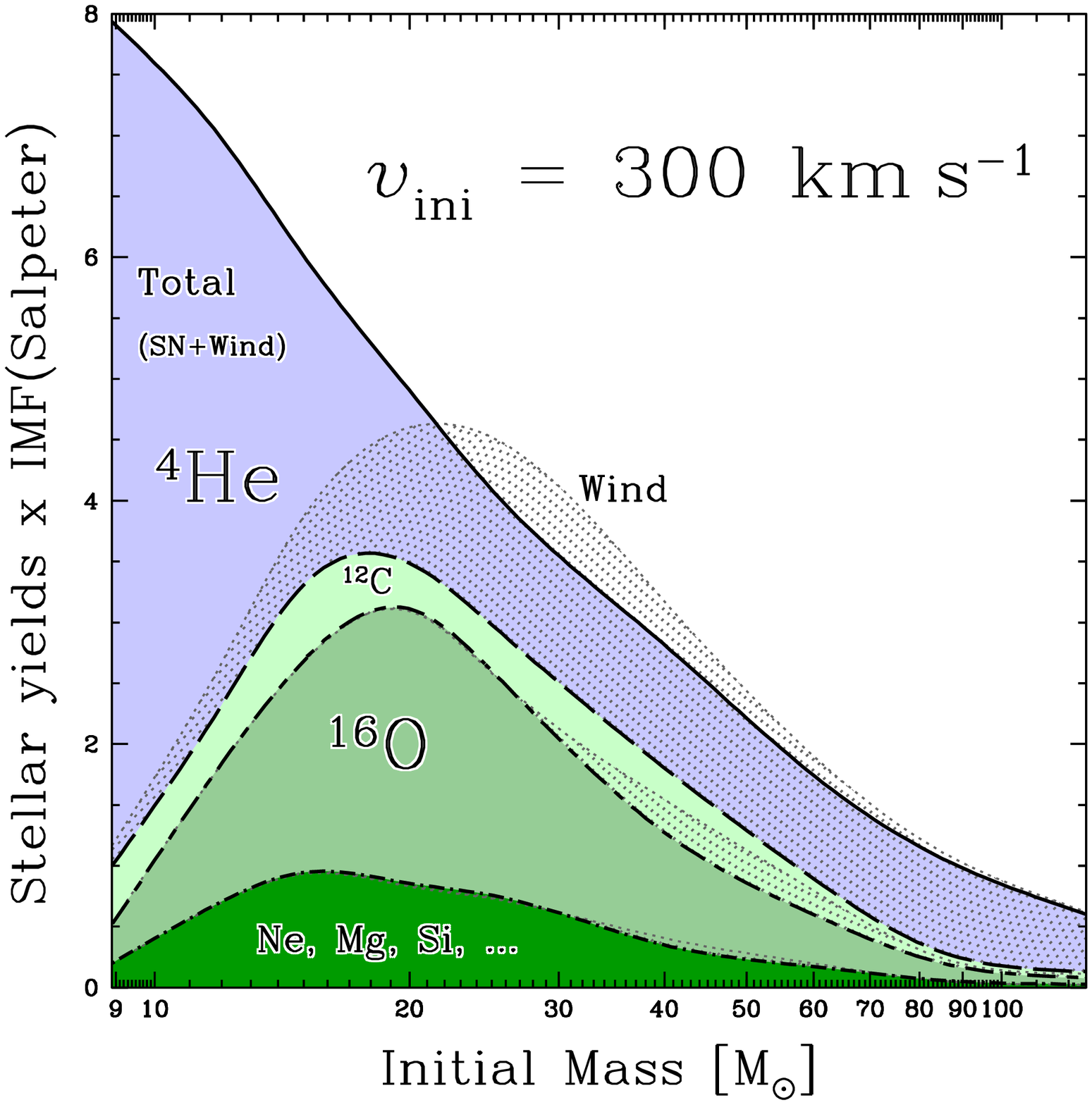}
\caption{{\it Left panel}: The yields x IMF for models without rotation. 
{\it Right panel}: The yields x IMF
for models with $\upsilon_{\mathrm{ini}}= 300$ km/s (Hirschi et al. 2005). }
\label{yhimf}
\end{figure}

Fig.~\ref{yh}
shows the chemical yields from models without and with rotation (Hirschi et al. 2005). Fig.~\ref{yhimf}
 shows these  yields multiplied by the initial mass function (IMF). The main
 conclusion is that below an initial mass of 30 $M_{\odot}$, the cores are larger 
 and thus the production of   $\alpha$--elements is enhanced. Typically a rotating 20 M$_\odot$
 model will have more or less the same nucleosynthetic contribution as a non-rotating 30 M$_\odot$
 stellar model.
 Above  30 $M_{\odot}$,
 mass loss is  the dominant effect and more He is ejected before being 
 further processed, while the size of the core is only slightly reduced.
 When we account for  the weighting by the  IMF, the production of oxygen and
 of $\alpha$--elements  is globally enhanced as illustrated by Fig.~\ref{yhimf}, while the effect on the He--production in massive stars remains limited. 

\subsection{At higher than solar metallicity}

\begin{figure}
\includegraphics[width=4.4in,height=2.5in]{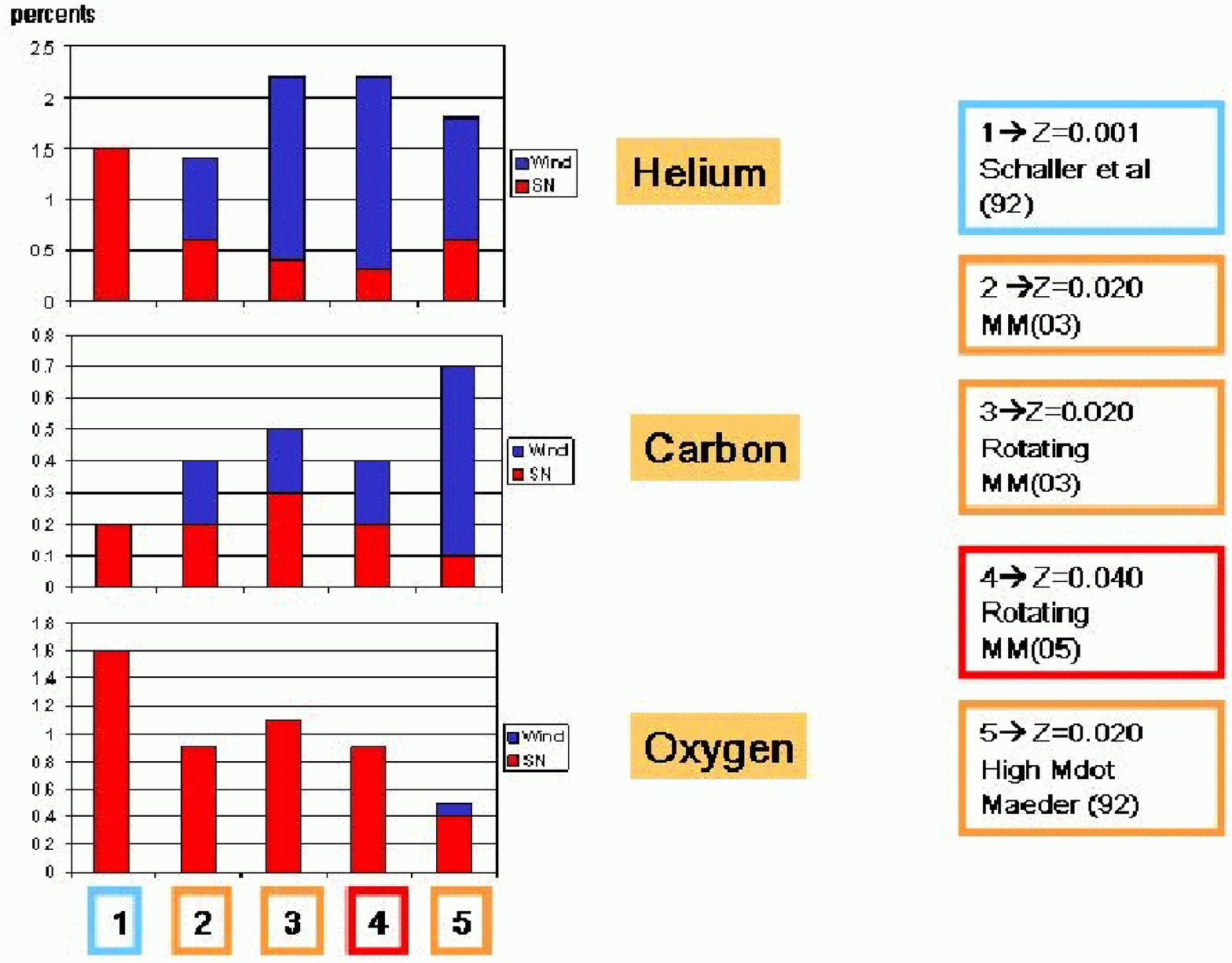}
\caption{Mass of new helium, carbon and oxygen ejected by stars more massive than 8 M$_\odot$
per mass initially in stars given by different stellar models. A Salpeter IMF has been used. The labels MM03 and MM05
are for respectively Meynet \& Maeder 2003 and Meynet \& Maeder 2005. }\label{fig7}
\end{figure}

In the high metallicity regime, mass loss is the dominant factor affecting both the evolution and the nucleosynthetic contributions of the massive stars. The yields
of  helium, carbon and oxygen are shown in Fig~\ref{fig7}. We can see the importance of mass loss for the carbon yields. Indeed the 
larger yields are obtained for the models computed with enhanced mass loss rates. It is interesting here to
note that high mass loss is not a sufficient condition for obtaining high carbon yields. High carbon yields are obtained only when the star enters into the WC phase at an early stage of the core He-burning phase.
This can be seen 
comparing the 60 M$_\odot$ stellar model at $Z=0.04$ computed by Meynet \& Maeder (2005) with the 60 M$_\odot$ stellar model at $Z=0.02$ with enhanced mass loss rate computed by Maeder (1992).
These two models end their lifetimes with similar final masses (the model of Meynet \& Maeder 2005 with 6.7 M$_\odot$ and the model of Maeder 1992 with 7.8 M$_\odot$), thus the mass of new carbon ejected at the time
of the supernova explosion is quite similar for both models and is around 0.45 M$_\odot$. When one
compares the mass of new carbon ejected by the winds, we obtain 7.3 and 0.16  M$_\odot$ for respectively the
model of Maeder (1992) and the one of Meynet \& Maeder (2005). This difference arises from the fact that the model
with enhanced mass loss enters the WC phase at a much earlier time of the core He-burning phase, typically when
the mass fraction of helium at the centre, $Y_c$ is 0.43 and the actual mass of the star is 25.8 M$_\odot$,
while the model of Meynet \& Maeder (2005) enters the WC stage when $Y_c$ is 0.24 and the actual mass 8.6 M$_\odot$. 

The example above shows that the entry at an early stage into the WC phase is more favored
by strong mass loss than by rotation. This comes from the fact that rotation favors an early entry into the
WN phase, while the star has still an important H-rich envelope. It takes time for the whole H-rich envelope to be removed and when it is done, the star is already well advanced into the core He-burning phase.
Of course this conclusion is quite dependent on the magnitudes of the mass loss rates. For instance higher
mass loss rates during (a part of) the WNL phase would favor an early entrance into the WC phase. This would give a better agreement with the number ratio of WC to WN observed and would lead to higher carbon yields.

Let us conclude this section by estimating an empirical yield in carbon from the Wolf-Rayet stars in the solar neighborhood. From the catalogue by van der Hucht (2001) one obtains that the number of WC stars at a distance less than 3 kpc around the Sun is 44. The mass loss rate during the WC phase is estimated to
be 10$^{-4.8}$ M$_\odot$ per year, the mass fraction of carbon observed in the WC stellar wind is around 0.35 (see Table 2 in the review by Crowther 2007). If we
consider a star formation rate of 2-4 M$_\odot$ per square parsec and per Gyr, one obtains that the mass of (new) carbon ejected by WC wind per mass used to form stars is between 0.25 and 0.5\%. These empirical yields are well in lines with the range of values given by the models 
which are between 0.2-0.6\% (see Fig.~\ref{fig7}, the greatest value corresponding to the case of Maeder 1992). Let us note that incorporating the yields of Maeder (1992) into chemical evolution models has an important impact (see e.g. Prantzos et al. 1996). If the upper value of the empirical yields is the correct one then this indicates that
WC stars are very important sources of carbon in metal rich regions.

\subsection{At lower than solar metallicity}

Effects of rotation in metal poor stellar models are very important and have many interesting
consequences for nucleosynthesis.
Some of these effects are mainly due to the more efficient mixing obtained in metal poor stars and do not much depend on the mass loss induced by rotation, others are consequences of both effects.
We tentatively propose to call ``{\it spinstars}'' those stars whose evolution is deeply affected by fast axial rotation. The term of spinstar is  equivalent to a fast rotating star on the MS phase. 

Among the effects mainly due to enhanced rotational mixing, let us mention the fact that massive spinstars might be very efficient sources of primary nitrogen in metal poor regions (Meynet \& Maeder 2002ab; Chiappini et al. 2005, 2006ab) and lead to different trends for C/O and N/O at very
low metallicity. 
Other isotopes such as $^{13}$C, $^{18}$O could also be produced abundantly in such models.

Interesting consequences resulting from both enhanced rotational mixing and mass loss are:
\begin{itemize}
\item the possibility to explain the origin of the peculiar abundance pattern exhibited by the extremely metal poor
C-rich stars. These stars could be formed from wind material of massive ``spinstars'' or from material ejected, either in a mass transfer episode or by winds, from an E-AGB whose progenitor was a ``spinstar'' (Meynet et al. 2006; Hirschi 2007).

\item to provide an explanation for the origin of the helium-rich stars in $\omega$ Cen. The presence of a blue
ZAMS sequence in this cluster  (in addition to a red sequence which is about a factor two less rich in iron) is interpreted as the existence in this cluster of very helium rich stars. Typically stars on the blue sequence would have, according to stellar models, a mass fraction of helium of 0.40, while stars on the red
sequence would only have an helium mass fraction of 0.25 (Norris 2004). We have proposed that the helium-rich stars could be formed from wind material of massive spinstars (Maeder \& Meynet 2006). This material would indeed have the appropriate chemical composition for accounting for the abundance patterns observed in the blue sequence.

\item Interestingly, massive spinstars, losing mass at the critical limit, could also contribute
in providing material for forming second generation stars  in globular clusters. Such stars
would present peculiar surface abundances, relics of their nuclear-processing in the fast
rotating massive stars (Decressin et al. 2007 and the paper by Charbonnel in the present volume).

\end{itemize}

\section{Conclusions}

It is now a well known fact that rotation is a key feature of the evolution of stars. For metallicities
between those of the Small Magellanic Cloud and of the solar neighborhood, rotating models better reproduce the observed characteristics of stars than non-rotating models (see e.g. the discussion in Meynet et al. 2006). From this argument alone, one would expect that
they would do the same in metal poor regions. Of course at very low metallicity, direct comparisons between
massive star models and observed stars are no longer possible. Thus it is particularly important in this case
to first check the models in metallicity range where such comparisons are possible. 
When the same physical processes as those necessary to obtain good fits at high metallicity are accounted for in fast rotating metal poor stars, the model stars are on one hand strongly mixed and on the other hand may lose large amounts of material. These features might be very helpful for explaining numerous puzzling observational facts
concerning metal poor stars as seen above. Moreover, such models stimulate new questions which can
be the subjects of future works:
 
\begin{itemize}
\item One can wonder what would be the contribution of very fast rotating Pop III massive stars to the ionizing flux. These stars would follow an homogeneous evolution, 
evolve in the blue side of the HR diagram, would have their lifetimes
increased and would become WR stars. For all these reasons, they would likely be important sources of ionizing photons.

\item What would be the ultimate fate of such fast rotating Pop III stars? Would they give birth to collapsars as proposed by Yoon \& Langer (2005) and Woosley \& Heger (2006)? 

\item If pair instability supernovae have left no nucleosynthetic sign of their existence in observed metal
poor star, is this
because no sufficiently high massive stars have ever formed? Or, if formed, might these stars have skipped the
Pair Instability regime due to strong mass loss triggered by fast rotation?

\item Does a dynamo mechanism work in Pop III stars? This mechanism needs a small
pristine magnetic field which will be amplified at the expense of the differential rotational energy.
But does this pristine magnetic field exist in this case? 
In case the mechanism is working, how does its effects vary as a function of the metallicity?

\item Could the first generations of massive stars be important producers of helium as was suspected
long time ago by Tayler \& Hoyle (1964)? 

\item What was the distribution of the rotational velocities at different metallicities? 
Is this distribution the same in the field and in dense clusters (like globular clusters)?
As recalled above, some indirect observations indicate that the distribution might be biased toward
fast rotators at low metallicity.  
Are there any other indirect hints supporting this view? What would be the
physical mechanism responsible for such a trend? (Shorter disk locking episode in metal poor regions?).

\end{itemize}

The list above is not exhaustive. It simply reflects the richness of the subject which will certainly become a very fruitful area of research in the coming years.

\begin{acknowledgements}
My warm thanks to Andr\'e Maeder whose enlightened theoretical developments allowed to explore
new effects of rotation in stellar models. My deep thanks also to Raphael Hirschi, Patrick Eggenberger
and Sylvia Ekstr\"om who contributed a lot in exploring the effects of rotations in different mass ranges, metallicities, stellar evolutionary phases and in the context of asteroseismology. 
I thank also Cyril Georgy for a careful reading of the manuscript.
\end{acknowledgements}

\end{document}